\documentclass{ws-esc}
\usepackage{txfonts}
\usepackage{multicol,float}
\usepackage{graphicx}
\usepackage{enumitem}
\usepackage{mathtools}
\usepackage{multirow}
\usepackage{verbatim}

\begin{document}

\catchline{1}{1}{2016}{article id}{}
\markboth{Chatla, Chen and Shmueli}{Selected Topics in Statistical Computing}

\title{Selected Topics in Statistical Computing}

\author{Suneel Babu Chatla $^\dagger$ 
, Chun-houh Chen $^\ddagger$, and Galit Shmueli$^\dagger$} 

\address{$^\dagger$Institute of Service Science, National Tsing Hua University, Hsinchu 30013, Taiwan R.O.C \\[1pt]
$^\ddagger$Institute of Statistical Science, Academia Sinica, Taipei 11529, Taiwan R.O.C\\[1pt]}

\maketitle

\footnotetext{$^\dagger$ The corresponding author can be reached at suneel.chatla@iss.nthu.edu.tw. The first and third authors were supported in part by grant 105-2410-H-007-034-MY3 from the Ministry of Science and Technology in Taiwan.
}

\begin{history}
\received{Day Month Year}
\revised{Day Month Year}
\accepted{Day Month Year}
\published{Day Month Year}
\end{history}

\begin{abstract}
The field of computational statistics refers to statistical methods or tools that are computationally intensive. Due to the recent advances in computing power some of these methods have become prominent and central to modern data analysis. In this article we focus on several of the main methods including density estimation, kernel smoothing, smoothing splines, and additive models. While the field of computational statistics includes many more methods, this article serves as a brief introduction to selected popular topics. 
\end{abstract}


\footnotetext{This is an Open Access article published by World Scientific Publishing Company. It is distributed under the terms of the Creative Commons Attribution 3.0 (CC-BY) License. Further distribution of this work is permitted, provided the original work is properly cited.}

\begin{multicols}{2}
\section*{Introduction}
\begin{flushright}
\textit{- ``Let the data speak for themselves"}
\end{flushright}

In 1962, John Tukey\cite{tuk62} published an article on "the future of data analysis", which turns out to be extraordinarily clairvoyant. Specifically, he accorded algorithmic models as the same foundation status as  algebraic models that statisticians had favored at that time.
More than a three decades later, in 1998 Jerome Friedman delivered a keynote speech\cite{fri98} in which he stressed the role of data driven or algorithmic models in the next revolution of statistical computing. In response, the field of statistics has seen tremendous growth in some research areas related to computational statistics. 

According to the current Wikipedia  entry on ``Computational Statistics"\footnote{\url{https://en.wikipedia.org/wiki/Computational_statistics}, accessed August 24, 2016}: ``Computational statistics or statistical computing refers to the interface between statistics and computer science. It is the area of computational science (or scientific computing) specific to the mathematical science of statistics." Two well known examples of statistical computing methods are the bootstrap and Markov Chain Monte Carlo (MCMC). These methods are prohibitive with insufficient computing power. While the bootstrap has gained significant popularity both in academic research and in practical applications its feasibility still relies on efficient computing. Similarly, MCMC, which is at the core of Bayesian analysis, is computationally very demanding. A third method which has also become prominent in both academia and practice is nonparametric estimation. Today, nonparametric models are popular data analytic tools due to their flexibility despite being very computationally intensive, and even prohibitively intensive with large datasets.

In this article, we provide summarized expositions for some of these important methods. The choice of methods highlights major computational methods for estimation and for inference. We do not aim to provide a comprehensive review of each of these methods, but rather a brief introduction. However, we compiled a list of references for readers interested in further information on any of these methods. 
For each of the methods, we provide the statistical definition and properties, as well as a brief illustration  using an example dataset.

In addition to the aforementioned topics, the twenty first century has witnessed tremendous growth in statistical computational methods such as functional data analysis, lasso, and machine learning methods such as random forests, neural networks, deep learning and support vector machines. Although most of these methods have roots in the machine learning field, they have become popular in the field of statistics as well. The recent book by Ref.~\refcite{efr16} describes many of these topics.

The article is organized as follows. In Section \ref{sec-density}, we open with nonparametric density estimation. Sections \ref{sec-kernel} and \ref{sec-splines} discuss smoothing methods and their extensions. Specifically, Section \ref{sec-kernel} focuses on kernel smoothing while Section \ref{sec-splines} introduces spline smoothing. 
Section \ref{sec-additive} covers additive models, and Section \ref{sec-mcmc} introduces Markov chain Monte Carlo methods (MCMC). The final Section \ref{sec-resampling} is dedicated to the two most popular resampling methods: the bootstrap and jackknife. 

\section{Density Estimation}\label{sec-density}
A basic characteristic describing the behavior of any random variable $X$ is its probability density function. Knowledge of the density function is useful in many aspects. By looking at the density function chart we can get a clear picture of whether the distribution is skewed, multi-modal, etc. In the simple case of a continuous random variable $X$ over an interval $X \in (a,b)$, the density is defined as
\begin{align*}
P(a<X<b) &= \int_a^b f(x)dx.
\end{align*}
In most practical studies the density of $X$ is not directly available. Instead, we are given a set of $n$ observations $x_1,\ldots ,x_n$ 
that we assume are iid realizations of the random variable. We then aim to estimate the density on the basis of these observations. There are two basic estimation approaches: the parametric approach, which consists of representing the density with a finite set of parameters, and the nonparametric approach, which does not restrict the possible form of the density function by assuming it belongs to a pre-specified family of density functions.  

In parametric estimation only the parameters are unknown. Hence, the density estimation problem is equivalent to estimating the parameters. However, in the nonparametric approach, one must estimate the entire distribution. This is because we make no assumptions about the density function. 

\subsection{Histogram}
The oldest and most widely used density estimator is the histogram. Detailed discussions are found in Refs.~\refcite{sim12} and \refcite{har12}. Using the definition of derivatives we can write the density in the following form:
\begin{align}
\label{deriv}
f(x) \equiv \frac{d}{dx}F(x) \equiv \underset{h \rightarrow 0}{lim} \frac{F(x+h)-F(x)}{h}.
\end{align}
where $F(x)$ is the cumulative distribution function of the random variable $X$. A natural finite sample analog of equation (\ref{deriv}) is to divide the  real line into $K$ equi-sized bins with small bin width $h$ and replace $F(x)$ with the empirical cumulative distribution function
\begin{align*}
\hat{F}(x) &= \frac{\#\{x_i \le x\}}{n}.
\end{align*}
This leads to the empirical density function estimate
\begin{align*}
\hat{f}(x) &= \frac{(\#\{x_i \le b_{j+1}\}-\#\{x_i \le b_j\})/n}{h}, \quad x \in (b_j,b_{j+1}],
\end{align*}
where $(b_j,b_{j+1}]$ defines the boundaries of the $j$th bin and $h=b_{j+1}-b_j$. If we define $n_j=\#\{x_i \le b_{j+1}\}-\#\{x_i \le b_j\}$ then
\begin{align}
\hat{f}(x) &= \frac{n_j}{nh}.
\end{align}

The same histogram estimate can also be obtained using maximum likelihood estimation methods. Here, we try to find a density $\hat{f}$ maximizing the likelihood in the observations
\begin{align}
\Pi_{i=1}^{n} \hat{f}(x_i).
\end{align}

Since the above likelihood (or its logarithm) cannot be maximized directly, penalized maximum likelihood estimation can be used to obtain the histogram estimate.

Next, we proceed to calculate the bias, variance and MSE of the histogram estimator. These properties give us an idea of the accuracy and precision of the estimator. If we define
\begin{align*}
B_j &=\Big[x_0+(j-1)h, x_0+jh \Big), \quad j \in \mathbb{Z},
\end{align*}
with $x_0$ being the origin of the histogram, then the histogram estimator can be formally written as 
\begin{align}
\hat{f}_h(x)  &= (nh)^{-1} \sum_{i=1}^n\sum_j I(X_i \in B_j) I(x \in B_j).
\end{align}

We now define the bias of histogram estimator. Assume that the origin of the histogram $x_0$ is zero and $x \in B_j$. Since $X_i$ are identically distributed
\begin{align*}
E[\hat{f}_h(x)] &= (nh)^{-1} \sum_{i=1}^n E[I(X_i \in B_j)] \\
&= (nh)^{-1} n E[I(X \in B_j)] \\
&= h^{-1} \int_{(j-1)h}^{jh} f(u)du.
\end{align*}
This last term is not equal to $f(x)$ unless $f(x)$ is constant in $B_j$. For simplicity, assume $f(x)=a+cx, x \in B_j$ and $a,c \in \mathbb{R}$. Therefore
\begin{align*}
Bias[\hat{f}_h(x)] &= E[\hat{f}_h(x)] - f(x) \\
 &= h^{-1} \int_{B_j} (f(u)-f(x))du \\
&= h^{-1} \int_{B_j} (a+cu-a-cx)du \\
&= h^{-1} h c \left(\left(j-\frac{1}{2}\right)h-x \right) \\
&= c \left(\left(j-\frac{1}{2}\right)h-x \right).
\end{align*}
Instead of slope $c$ we may write the first derivative of the density at the midpoint $(j-\frac{1}{2})h$ of the bin $B_j$
\begin{align*}
Bias (\hat{f}_h(x)) &= f'\left(\left(j-\frac{1}{2}\right)h\right) \left(\left(j-\frac{1}{2}\right)h-x \right) \\
 &= O(1) O(h) \\
  &= O(h), \quad h \rightarrow 0.
\end{align*}
When $f$ is not linear, a Taylor expansion of $f$ to the first order reduces the problem to the linear case. Hence the bias of the histogram is given by
\begin{align}
Bias (\hat{f}_h(x)) &= \left(\left(j-\frac{1}{2}\right)h-x \right) f'\left(\left(j-\frac{1}{2}\right)h\right) + o(h) , \quad h \rightarrow 0.
\end{align}
Similarly, the variance for the histogram estimator can be calculated as
\begin{align*}
Var(\hat{f}_h(x)) &= Var\bigg((nh)^{-1} \sum_{i=1}^n I(X_i \in B_j) \bigg) \\
&= (nh)^{-2}\sum_{i=1}^n Var\Big[I(X_i \in B_j)\Big] \\
&= n^{-1}h^{-2} Var\Big[I(X \in B_j)\Big] \\
&= n^{-1}h^{-2} (\int_{B_j} f(u) du)(1-\int_{B_j} f(u) du) \\
&= (nh)^{-1} (h^{-1}\int_{B_j} f(u)du)(1-O(h)) \\
&= (nh)^{-1} (f(x)+o(1)), \quad h \rightarrow 0, nh\rightarrow \infty.
\end{align*}

\begin{figure}[H]
\centerline{
\includegraphics[width=8cm]{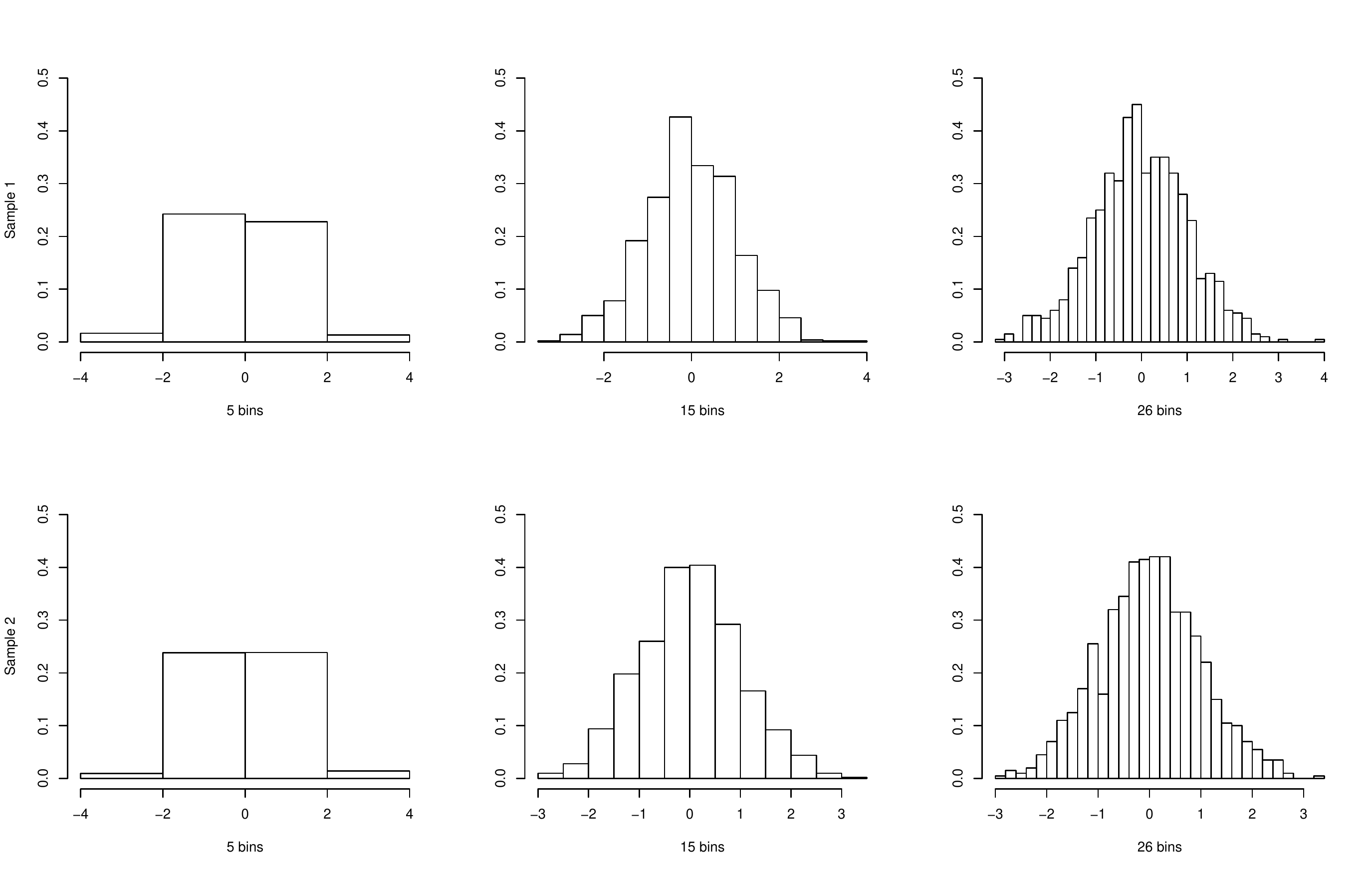}}
\caption{Histograms for two randomly simulated normal samples with 5 bins (left), 15 bins (middle), and 26 bins (right).}
\label{hbv}
\end{figure}

Bin width choice is crucial in constructing a histogram. As illustrated in Figure \ref{hbv}, bandwidth choice affects the bias-variance trade-off. The top three represent the histograms for the normal random sample but with three different bin sizes. Similarly bottom three histograms are from another normal sample. From the plot it can be seen that the histograms with larger bin width have smaller variability but larger bias and vice versa. Hence we need to strike a balance between bias and variance to come up with a good histogram estimator.

We observe that the variance of the histogram is proportional to $f(x)$ and decreases as $nh$ increases. This contradicts with the fact that the bias of the histogram decreases as $h$ decreases. To find a compromise we consider the mean square error (MSE):
\begin{align*}
MSE(\hat{f}_h(x)) &= Var(\hat{f}_h(x))+(Bias(\hat{f}_h(x)))^2 \\
&= \frac{1}{nh}f(x)+((j-1/2)h-x)^2f'((j-1/2)h)^2\\&+o(h)+o(\frac{1}{nh}).
\end{align*}
In order for the histogram estimator to be consistent, the $MSE$ should converges to zero asymptotically. Which means that the bin width should get smaller with the number of observations per bin $n_j$ getting larger as $n \rightarrow \infty$. 
Thus, under $nh \rightarrow \infty, h\rightarrow 0$, the histogram estimator is consistent; $\hat{f}_h(x) \xrightarrow{p}f(x)$.

Implementing the MSE using the formula is difficult in practice because of the unknown density involved. In addition, it should be calculated for each and every point. Instead of looking at the estimate at one particular point it might be worth calculating a measure of goodness of fit for the entire histogram. For this reason the mean integrated squared error (MISE) is used. It is defined as:
\begin{align*}
MISE(\hat{f}_h(x)) &= E\Bigg[\int_{-\infty}^{\infty} (\hat{f}-f)^2(x) dx \Bigg] \\
&= \int_{-\infty}^{\infty} MSE(\hat{f}_h(x)) dx) \\
&= (nh)^{-1}+h^2/12 ||f'||_2^2+o(h^2)+o((nh)^{-1}).
\end{align*}
Note that $||f'||_2^2$ ($dx$ is omitted in shorthand notation)is the square of the $L_2$ norm of $f'$ which describes how smooth the density function $f$ is. The common approach for minimizing MISE is to minimize it as a function of $h$ without higher order terms (Asymptotic MISE, or AMISE). The minimizer ($h_0$), called an optimal bandwidth, can be obtained by differentiating AMISE with respect to $h$.
\begin{align}
h_0 =\bigg( \frac{6}{n||f'||_2^2}\bigg).
\label{obw}
\end{align}
Hence we see that for minimizing AMISE we should theoretically choose $h_0 \sim n^{-1/3}$, which if we substitute in the MISE formula, would give the best convergence rate $O(n^{-2/3})$ for a sufficiently large $n$. Again, the solution of equation \ref{obw} does not help much as it involves $f'$ which is still unknown. However, this problem can be overcome by using any reference distribution (e.g., Gaussian). This method is often called the ``plug-in" method.

\subsection{Kernel Density Estimation}
\begin{figure}[H]
\centerline{
\includegraphics[width=8cm]{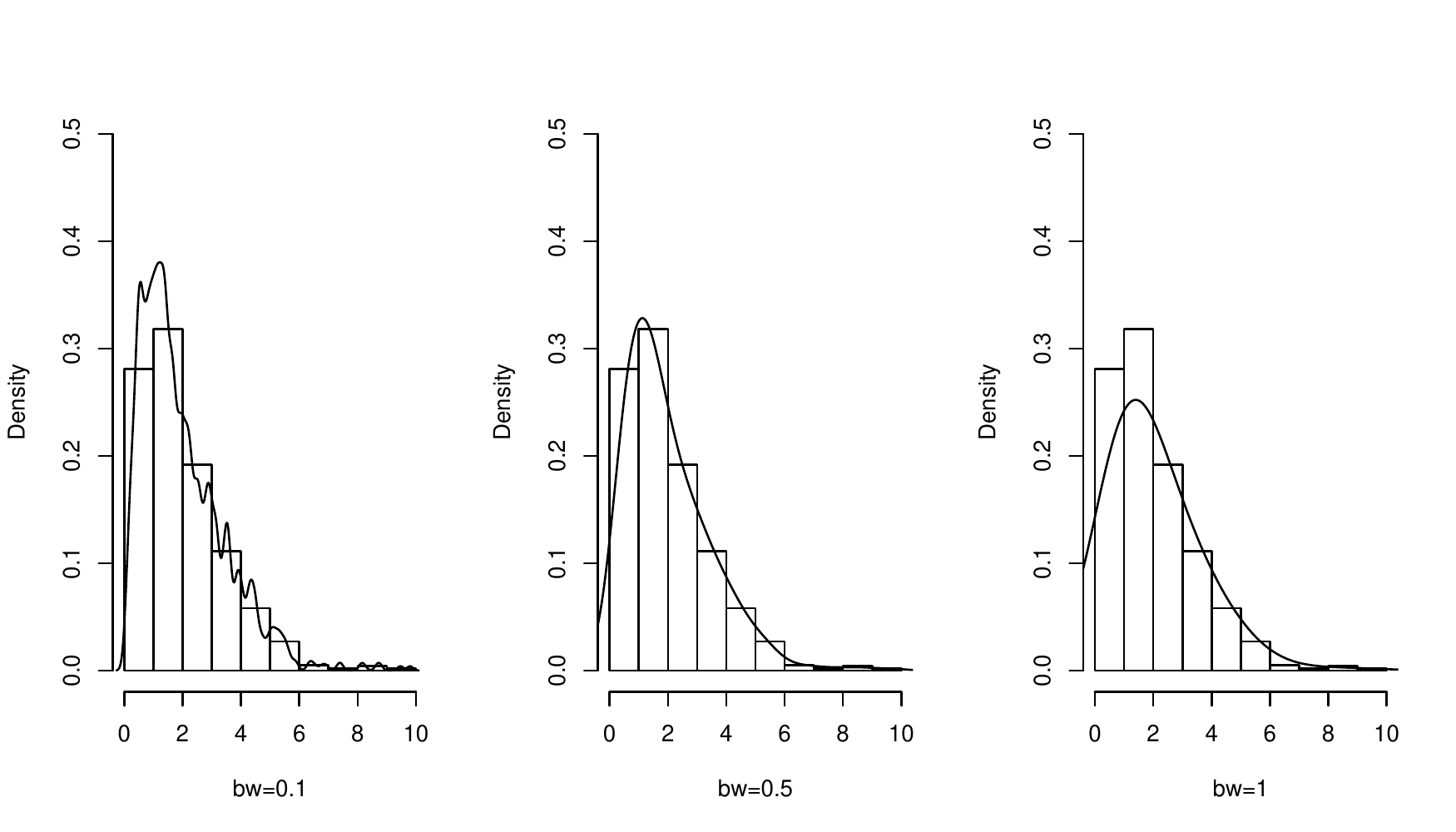}}
\caption{Kernel densities with three bandwidth choices (0.1, 0.5, and 1) for a sample from an exponential distribution.}
\label{kbv}
\end{figure}

The idea of the kernel estimator was introduced by Ref.~\refcite{ros56}. 
Using the definition of the probability density, suppose $X$ has density $f$. Then
\begin{align*}
f(x) &= \underset{h \rightarrow 0}{lim} \frac{1}{2h} P(x-h<X<x+h).
\end{align*}
For any given $h$, we can estimate the probability $P(x-h<X<x+h)$ by the proportion of the observations falling in the interval $(x-h,x+h)$. Thus a naive estimator $\hat{f}$ of the density is given by 
\begin{align*}
\hat{f}(x) &= \frac{1}{2hn}\sum_{i=1}^n I_{(x-h,x+h)}(X_i).
\end{align*}
To express the estimator more formally, we define the weight function $w$ by
\begin{align*}
w(x) &= \begin{cases} \frac{1}{2} & \text{if} \quad  |x|<1 \\
0 & \text{otherwise}.
\end{cases}
\end{align*}
Then it is easy to see that the above naive estimator can be written as 
\begin{align}
\hat{f}(x) &= \frac{1}{n}\sum_{i=1}^n \frac{1}{h}w\bigg(\frac{x-X_i}{h}\bigg).
\end{align}
However, the naive estimator is not wholly satisfactory because $\hat{f}(x)$ is of a ``stepwise" nature and not differentiable everywhere. We therefore generalize the naive estimator to overcome some of these difficulties by replacing the weight function $w$ with a kernel function $K$ which satisfies the conditions
\begin{align*}
\int K(t) dt =1, \int tK(t)=0,\quad \text{and} \quad \int t^2K(t)dt=k_2 \neq 0.
\end{align*}
Usually, but not always, $K$ will be a symmetric probability density function. Now the kernel density estimator  becomes
\begin{align*}
\hat{f}(x) &= \frac{1}{nh}\sum_{i=1}^{n}K\bigg(\frac{x-X_i}{h}\bigg).
\end{align*}
From the kernel density definition it can be observed that
\begin{itemize}
\item Kernel functions are symmetric around 0 and integrate to 1
\item Since the kernel is a density function, the kernel estimate is a density too: 
$\int K(x)dx=1$ implies $\int \hat{f}_h(x)dx=1$.
\item The property of smoothness of kernels is inherited by $\hat{f}_h(x)$. If $K$ is $n$ times continuously differentiable, then $\hat{f}_h(x)$ is also $n$ times continuously differentiable.
\item Unlike histograms, kernel estimates do not depend on the choice of origin.
\item Usually kernels are positive to assure that $\hat{f}_h(x)$ is a density. There are reasons to consider negative kernels 
but then $\hat{f}_h(x)$ may be sometimes negative.
\end{itemize}
We next consider the bias of the kernel estimate:
\begin{align*}
Bias[\hat{f}_h(x)] &= E[\hat{f}_h(x)] - f(x) \\
&= \int K(s) f(x+sh)ds -f(x) \\
&= \int K(s) [f(x)+shf'(x)+\frac{h^2s^2}{2}f''(x) \\&+o(h^2)] ds -f(x)\\
&= \frac{h^2}{2} f''(x)k_2+o(h^2), \quad h \rightarrow 0.
\end{align*}
For the proof see Ref.~\refcite{pur62}. We see that the bias is quadratic in $h$. Hence we must choose $h$ small to reduce the bias.
Similarly the variance for the kernel estimate can be written as
\begin{align*}
var(\hat{f}_h(x)) &= n^{-2} var\bigg(\sum_{i=1}^n K_h(x-X_i) \bigg) \\
&=n^{-1} var[K_h(x-X)] \\
&=(nh)^{-1} f(x) \int K^2  +o((nh)^{-1}), nh \rightarrow \infty.
\end{align*}
Similar to the histogram case, we observe a bias-variance tradeoff. The variance is nearly proportional to $(nh)^{-1}$, which requires choosing $h$ large for minimizing variance. However, this contradicts with the aim of decreasing bias by choosing small $h$. From a smoothing perspective, smaller bandwidth results in under-smoothing  and larger bandwidth results in over-smoothing. From the illustration in Figure \ref{kbv}, we see that when the bandwidth is too small (left) the kernel estimator under-smoothes the true density and when the bandwidth is large (right) the kernel estimator over-smoothes the underlying density. Therefore we consider MISE or MSE of $h$ as a compromise.
\begin{align*}
MSE[\hat{f}_h(x)] &= \frac{1}{nh}f(x) \int K^2 +\frac{h^4}{4}(f''(x)k_2)^2+o((nh)^{-1})\\&+o(h^4), h\rightarrow 0, nh \rightarrow\infty.
\end{align*}
Note that $MSE[\hat{f}_h(x)]$ converges to zero, if $h \rightarrow 0$ and $nh \rightarrow \infty$. Thus the kernel density estimate is consistent, that is $\hat{f}_h(x) \xrightarrow{p} f(x)$. On the whole, the variance term in MSE penalizes under smoothing and the bias term penalizes over smoothing. 
Further, the asymptotic optimal bandwidth can be obtained by differentiating MSE with respect to $h$ and equating it to zero:
\begin{align*}
h_0 &= \bigg(\frac{\int K^2}{(f''(x))^2k_2^2n} \bigg)^{1/5}  .
\end{align*}
It can be further verified that if we substitute this bandwidth in the MISE formula then
\begin{align*}
MISE (\hat{f}_{h_0}) &= \frac{5}{4}\Big(\int K^2 \Big)^{4/5}k_2^{2/5} \Big(\int f''(x)^2\Big)^{1/5} n^{-4/5} \\
&= \frac{5}{4} C(K) \Big(\int f''(x)^2\Big)^{1/5} n^{-4/5}, \text{where} \\
  C(K) &= \Big(\int K^2 \Big)^{4/5}k_2^{2/5}.
\end{align*}
From the above formula it can be observed that we should choose a kernel $K$ with a small value of $C(K)$, when all other things are equal. The problem of minimizing $C(K)$ can be reduced to that of minimizing  $\int K^2$ by allowing suitable rescaled version of kernels. In a different context, Ref.~\refcite{hod56} showed that this problem can be solved by setting $K$ to be a \textit{Epanechnikov kernel}\cite{epa69} (see Table \ref{kernel}). 

We define the efficiency of any symmetric kernel $K$ by comparing it to the Epanechnikov kernel:
\begin{align*}
eff(K) &= {C(K_e)/C(K)}^{5/4} \\
& = \frac{3}{5\sqrt{5}} k_2^{-1/2} {\int K^2 }^{-1}.
\end{align*}
The reason for the power $5/4$ in the above equation is that for large $n$, the MISE will be the same as whether we use $n$ observations with kernel $K$ or $n$eff$(K)$ observations and the kernel $K_e$\cite{stu73}. Some kernels and their efficiencies are given in the Table \ref{kernel}.
\begin{table}[H]
\tbl{Definitions of some kernels and their efficiencies.}  
{\begin{tabular}{lll}
\toprule
Kernel & $K(t)$ & Efficiency \\
\colrule
Rectangular & $\begin{cases} \frac{1}{2} & \mbox{for} |t| \le 1,\\ 0 & \mbox{otherwise.} \end{cases}$ & $0.9295$\\ \\
Epanechnikov & $\begin{cases} \frac{3}{4}(1-\frac{1}{5}t^2)/\sqrt{5} & \mbox{for }  |t| \le 5,\\
  0  &  \mbox{otherwise.} \end{cases} $ & $1$ \\  \\
Biweight & $ \begin{cases} \frac{15}{16}(1-t^2)^2 & \mbox{for } |t|\le 1, \\
0 & \mbox{otherwise.} \end{cases} $ & $0.9939$ \\ \\
Triweight & $\begin{cases} \frac{35}{32} (1-t^2)^3 & \mbox{for } |t|\le 1, \\ 0 & \mbox{otherwise.} \end{cases}$ & $0.987$\\ \\
Triangular & $ \begin{cases} 1-|t| & \mbox{for } |t| \le 1, \\ 
0 & \mbox{otherwise.} \end{cases}$ & $0.9859$\\ \\
Gaussian & $\frac{1}{\sqrt{2\pi}} e^{-(1/2)t^2} $ & $0.9512$ \\ \\
\botrule
\end{tabular}} \label{kernel}
\end{table}
The top four kernels are particular cases of the following family:
\begin{align*}
K(x;p) &=\{2^{2p+1} B(p+1,p+1 \}^{-1}(1-x^2)^p I_{\{|x|< 1\}},
\end{align*}
where $B(\cdot,\cdot)$ is the beta function. These kernels are symmetric beta densities on the interval $[-1,1]$. For $p=0$ the expression gives rise to a rectangular density, $p=1$ to Epanechnikov, and $p=2$ and $p=3$ are bivariate and trivariate kernels, respectively. The standard normal density is obtained as the limiting case $p \rightarrow \infty$.

Similar to the histogram scenario, the problem of choosing the bandwidth (smoothing parameter) is of crucial importance to density estimation. A natural method for choosing the bandwidth is to plot several curves and choose the estimate that is most desirable. For many applications this approach is satisfactory. However, there is a need for data-driven and automatic procedures that are practical and have fast convergence rates. The problem of bandwidth selection has stimulated much research in kernel density estimation. The main approaches include cross-validation and ``plug-in" methods (see the review paper by Ref.~\refcite{jon96}). 

As an example, consider least squares cross validation, which was suggested by Ref.~\refcite{rud82} and \refcite{bow84} - see also Ref.~\refcite{bow84cross},\refcite{hal83} and \refcite{sto84}. Given any estimator $\hat{f}$ of a density $f$, the integrated squared error can be written as
\begin{align*}
\int (\hat{f}-f)^2 &= \int \hat{f}^2-2\int \hat{f}f+\int f^2.
\end{align*}
Since the last term ($\int f^2$) does not involve $f'$, the updated quantity $R(\hat{f})$ of the above equation, is 
\begin{align*}
R(\hat{f}) &= \int \hat{f}^2 -2\int \hat{f}f.
\end{align*}
The basic principle of least squares cross validation is to construct an estimate of $R(\hat{f})$ from the data themselves and then to minimize this estimate over $h$ to give the choice of window width. The term $\int \hat{f}^2$ can be found from the estimate $\hat{f}$. Further, if we define $\hat{f}_{-i}$ as the density estimate constructed from all the data points except $X_i$, then $\int \hat{f}f$ can be computed using
\begin{align*}
\hat{f}_{-i}(x) &=(n-1)^{-1}h^{-1} \sum_{j\neq i} K{h^{-1}(x-X_j)}.
\end{align*}
Now we define the required quantity without any unknown terms $f$, as
\begin{align*}
M_0(h) &= \int \hat{f}^2 -2n^{-1}\sum_{i}\hat{f}_{-i}(X_i) .
\end{align*}
The score $M_0$ depends only on data and the idea of least squares cross validation is to minimize the score over $h$. There also exists a computationally simple approach to estimate $M_0$ - Ref.~\refcite{sto84} provided the large sample properties for this estimator. Thus, asymptotically, least squares cross validation achieves the best possible choice of smoothing parameter, in the sense of minimizing the integrated squared error. For further details see Ref.~\refcite{sil86}. 

Another possible approach related to ``plug-in" estimators is to use a standard family of distributions as a reference for the value $\int f''(x)^2 dx$ which is the only unknown in the optimal bandwidth formula $h_{opt}$. For example, if we consider the normal distribution with variance $\sigma^2$, then  
\begin{align*}
\int f''(x)^2 dx = \sigma^{-5} \int \phi''(x)^2dx = \frac{3}{8} \pi^{-1/2}\sigma^{-5} \approx 0.212 \sigma^{-5}.
\end{align*}
If a Gaussian kernel is used, then the optimal bandwidth is
\begin{align*}
h_{opt} &= (4\pi)^{-1/10}\frac{3}{8}\pi^{-1/2}\sigma n^{-1/5}\\
&= \left(\frac{4}{3}\right)^{1/5}\sigma n^{-1/5} \\
&= 1.06 \sigma n^{-1/5}.
\end{align*}
A quick way of choosing the bandwidth is therefore estimating $\sigma$ from the data and substituting it in the above formula. While this works well if the normal distribution is a reasonable approximation, it may oversmooth if the population is multimodal. Better results can be obtained using a robust measure of spread such as the interquartile $(R)$ range, which in this example yields $h_{opt}=0.79Rn^{-1/5}$. Similarly, one can improve this further by taking the minimum of the standard deviation and the interquartile range divided by 1.34. For most applications these bandwidths are easy to evaluate and serve as a good starting value.

Although the underlying idea is very simple and the first paper was published long ago, the kernel approach did not make much progress until recently, with advances in computing power. At present, without an efficient algorithm, the calculation of a kernel density for moderately large datasets can become prohibitive. The direct use of the above formulas for computations is very inefficient. Researchers developed fast and efficient Fourier transformation methods to calculate the estimate using the fact that the kernel estimate can be written as a convolution of data and the kernel function \cite{loa06}.

\section{Kernel Smoothing}\label{sec-kernel}
The problem of smoothing sequences of observations is important in many branches of science and it is demonstrated by the number of different fields in which smoothing methods have been applied. Early contributions were made in fields as diverse as astronomy, actuarial science, and economics. Despite their long history, local regression methods have received little attention in the statistics literature until the late 1970s. Initial work includes the mathematical development of Refs.~\refcite{sto77}, \refcite{kat79} and \refcite{sto80}, and the LOWESS procedure of Ref.~\refcite{cle79}. Recent work on local regression includes Refs.~\refcite{fan92}, \refcite{fan93} and \refcite{has93}.

The local linear regression method was developed largely as an extension of parametric regression methods and accompanied by an elegant finite sample theory of linear estimation methods that build on theoretical results for parametric regression. It is a method for curve estimation by fitting locally weighted least squares regression. One extension of local linear regression, called local polynomial regression, is discussed in \cite{rup94} and in the monograph by Ref.~\refcite{fan96}. 

Assume that $(X_1,Y_1), \ldots,(X_n,Y_n)$ are iid observations with conditional mean and conditional variance  denoted respectively by
\begin{align}
m(x) = E(Y|X=x) \quad \text{and} \quad \sigma^2(x)=Var(Y|X=x).
\end{align}
Many important applications involve estimation of the regression function $m(x)$ or its $\nu^{th}$ derivative $m^{(\nu)}(x)$. The performance of an estimator $\hat{m}_{\nu}(x)$ of $m^{(\nu)}(x)$ is assessed via its MSE or MISE defined in previous sections. 
While the MSE criterion is used when the main objective is to estimate the function at the point $x$, MISE criterion is used when the main goal is to recover the whole curve. 

\subsection{Nadaraya-Watson Estimator}
If we do not assume a specific form for the regression function $m(x)$, then a data point remote from $x$ carries little information about the value of $m(x)$. In such a case, an intuitive estimator of the conditional mean function is the running locally weighted average. If we consider a kernel $K$ with bandwidth $h$ as the weight function, the Nadaraya-Watson kernel regression estimator is given by
\begin{align}
\hat{m}_h(x) &= \frac{\sum_{i=1}^n K_h(X_i-x)Y_i}{\sum_{i=1}^n K_h(X_i-x)},
\end{align}
where $K_h(\cdot)=K(\cdot/h)/h$. For more details see Refs.~\refcite{nad64},\refcite{wat64} and \refcite{har90}. 

\subsection{Gasser-M\"{u}ller Estimator}
Assume that the data have already been sorted according to the $X$ variable. Ref.~\refcite{gas79} proposed the following estimator:
\begin{align}
\hat{m}_h(x) &= \sum_{i=1}^n \int_{s_{i-1}}^{s_i} K_h(u-x)du Y_i,
\label{gmest}
\end{align}
with $s_i=(X_i+X_{i+1})/2, X_0=-\infty$ and $X_{n+1}=+\infty$. The weights in equation (\ref{gmest}) add up to 1, so there is no need for a denominator as in the Nadaraya-Watson estimator. Although it was originally developed for equispaced designs, the Gasser-M\"{u}ller estimator can also be used for non-equispaced designs. For the asymptotic properties please refer to Refs.~\refcite{mac89} and \refcite{chu91}.

\subsection{Local Linear Estimator}
\begin{figure}[H]
\centerline{
\includegraphics[width=8cm]{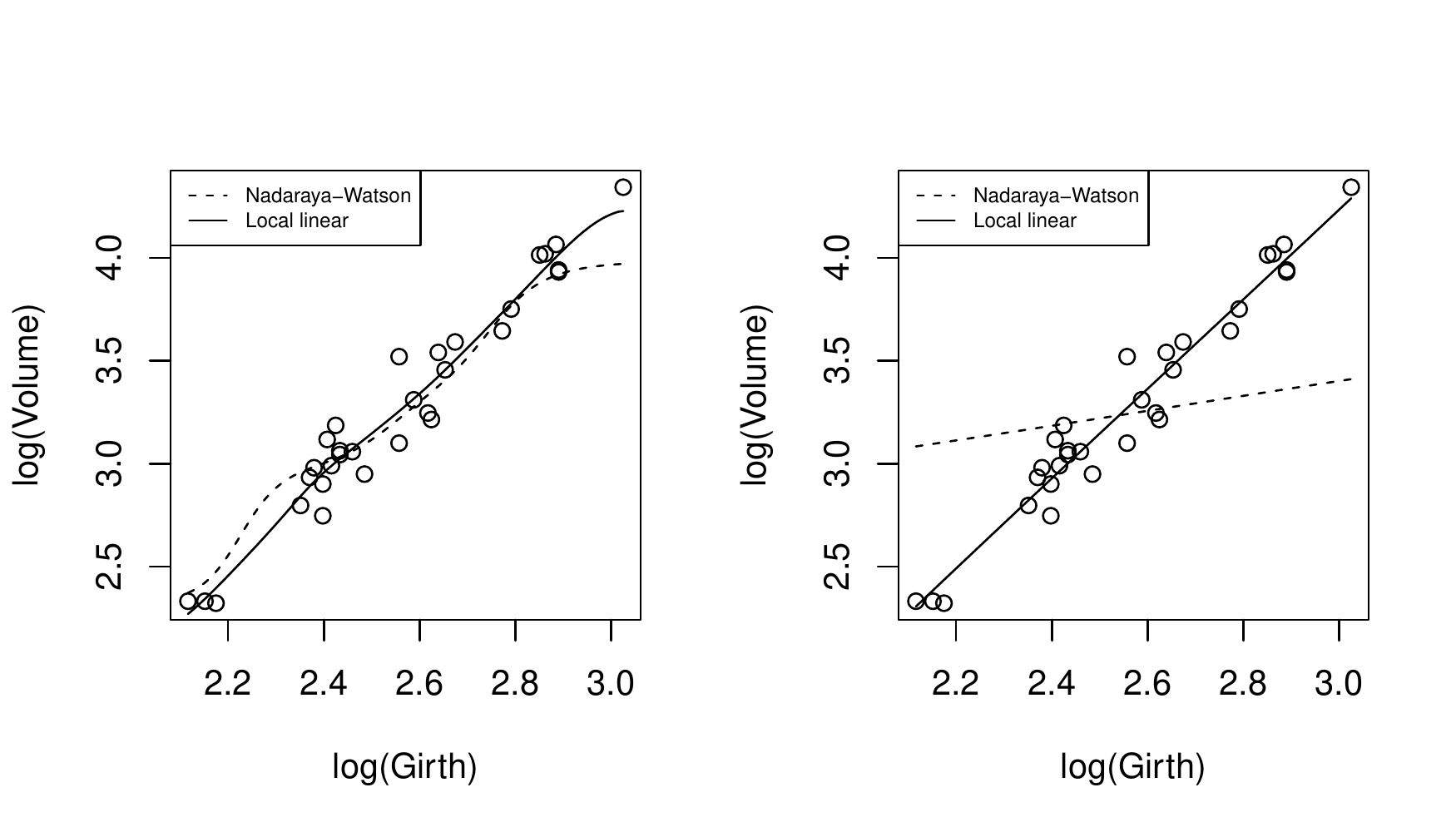}}
\caption{Comparison of local linear fit versus Nadaraya-Watson fit for a model of (log) volume as a function of (log) girth, with choice of two bandwidths: 0.1 (left) and 0.5 (right).}
\label{locplot}
\end{figure}

This estimator assumes that locally the regression function $m$ can be approximated by
\begin{align}
m(z) \approx \sum_{j=0}^p \frac{m^{(j)}(x)}{j!} (z-x)^j \equiv \sum_{j=0}^p\beta_j(z-x)^j,
\end{align}
for $z$ in a neighborhood of $x$, by using a Taylor's expansion. Using a local least squares formulation, the model coefficients can be estimated by minimizing the following function:
\begin{align}
\sum_{i=1}^{n} \left\{ Y_i - \sum_{j=0}^{p} \beta_j (X_i-x)^j \}^2 K_h (X_i-x) \right\},
\label{lleast}
\end{align}
where $K(\cdot)$ is a kernel function with bandwidth $h$. If we let $\hat{\beta}_j$ ($j=0,\cdots,p$) be equal to the estimates obtained from minimizing equation (\ref{lleast}), then the estimator for the regression functions are obtained as
\begin{align}
\hat{m}_v(x) &= v! \hat{\beta}_v.
\end{align}
When $p=1$ the estimator $\hat{m}_0(x)$ is termed a \emph{local linear smoother} or \emph{local linear estimator} with the following explicit expression:
\begin{align}
\hat{m}_0(x) = \frac{\sum_{1}^{n} w_iY_i}{\sum_{1}^{n} w_i}, w_i=K_h(X_i-x)\{S_{n,2}-(X_i-x)S_{n,1}\}.
\end{align}
where $S_{n,j}=\sum_1^n K_h(X_i-x)(X_i-x)^j$.
When $p=0$, the local linear estimator is equals to the Nadaraya-Watson estimator. Also, both Nadaraya-Watson and Gasser-M\"{u}ller estimators are of the type of local least squares estimator with weights $w_i=K_h(X_i-x)$ and $w_i=\int_{s{i-1}}^{s_i}K_h(u-x)du$, respectively. The asymptotic results are provided in Table \ref{locasym}, which is  taken from Ref.~\refcite{fan92}.
\begin{table}[H]
\tbl{Comparison of asymptotic properties of local estimators.}  
{\begin{tabular}{@{}lcc@{}}
\toprule
Method & Bias & Variance \\
\colrule
Nadaraya-Watson & $(m''(x)+\frac{2m'(x)f'(x)}{f(x)})b_n$ & $V_n$ \\
Gasser-M\"{u}ller & $m''(x)b_n$ & $1.5V_n$ \\
Local linear & $m''(x)b_n$ & $V_n$ \\
\botrule
\end{tabular}}
\begin{tabnote}
Here, $b_n=\frac{1}{2}\int_{-\infty}^{\infty}u^2 K(u)du h^2$ and $V_n=\frac{\sigma^2(x)}{f(x)nh}\int_{-\infty}^{\infty} K^2(u)du$.
\end{tabnote}
\label{locasym}
\end{table}
To illustrate both Nadaraya-Watson and local linear fit on data, we considered an example of a dataset on trees\cite{atk85,rya76}. This dataset includes measurements of the girth, height, and volume of timber in 31 felled back cherry trees. The smooth fit results are described in Figure \ref{locplot}. The fit results are produced using the \emph{KernSmooth}\cite{wan15} package in R-Software\cite{tea13}. In the right panel of Figure \ref{locplot}, we see that for a larger bandwidth, as expected, Nadaraya-Watson fits a global constant model while the local linear fits a linear model. Further, from the left panel, where we used a reasonable bandwidth, the Nadara-Watson estimator has more bias than the local linear fit.
In comparison with the local linear fit, the Nadaraya-Watson estimator locally uses one parameter less without reducing the asymptotic variance. It suffers from large bias, particularly in the region where the derivative of the regression function or design density is large. Also it does not adapt to non uniform designs. In addition, it was shown that the Nadaraya-Watson estimator has zero minimax efficiency - for details see Ref.~\refcite{fan92}. Based on the definition of minimax efficiency, a 90\% efficient estimator uses only 90\% of the data. Which means that the Nadaraya-Watson does not use all the available data.  
In contrast, the Gasser-M\"{u}ller estimator corrects the bias of the Nadaraya-Watson estimator but at the expense of increasing variability for for random designs. Further, both the Nadaraya-Watson and Gasser-M\"{u}ller estimator have a large order of bias when estimating a curve at the boundary region. Comparisons between local linear and local constant (Nadaraya-Watson) fit were discussed in detail by Refs.~\refcite{chu91,fan92} and \refcite{has93}.

\subsection{Computational Considerations}
Recent proposals for fast implementations of nonparametric curve estimators include the binning methods and the updating methods. Ref.~\refcite{fan94} gave careful speed comparisons of these two fast implementations and direct naive implementations under a variety of settings and using various machines and software. Both fast methods turned out to be much faster with negligible differences in accuracy. 

While the key idea of the binning method is to bin the data and compute the required quantities based on the binned data, the key idea of the updating method involves updating the quantities previously computed. It has been reported that for practical purposes neither method dominates the other.

\section{Smoothing Using Splines}\label{sec-splines}
Similar to local linear estimators, another family of methods that provide flexible data modeling is spline methods. These methods involve fitting piecewise polynomials or splines to allow the regression function to have discontinuities at certain locations which are called ``knots" \cite{eub88,wah90,gre94}.

\subsection{Polynomial Spline}
Suppose that we want to approximate the unknown regression function $m$ by a cubic spline function, that is, a piecewise polynomial with continuous first two derivatives. Let $t_1,\cdots,t_J$ be a fixed knot sequence such that $-\infty <t_1<\cdots <t_J<+\infty$. Then the cubic spline functions are twice continuously differentiable functions $s$ such  that restrictions of $s$ to each of the intervals $(-\infty,t_1],[t_1,t_2],\cdots,[t_{J-1},t_J],[t_J,+\infty)$ is a cubic polynomial. The collection of all these cubic spline functions forms a $(J+4)$-dimensional linear space. There exist two popular cubic spline bases for this linear space:
\begin{description}
\item[Power basis:] $1,x,x^2,x^3,(x-t_j)_+^3,(j=1,\cdots,J)$
\item[B-spline basis:] The $i$th B-spline of degree $p=3$, written as $N_{i,p}(u)$, is defined recursively as:
\begin{align*}
N_{i,0}(u) &= \begin{cases}
1 & \text{if} u_i \leq u \leq u_{i=1} \\
0 & \text{otherwise}.
\end{cases} \\
N_{i,p}(u) &= \frac{u-u_i}{u_{i+p}-u_i} N_i,{p-1}(u) + \frac{u_{i+p+1}-u}{u_{i+p+1}-u{i+1}} N_{i+1,p-1}(u).
\end{align*}
The above is usually referred to as the \emph{Cox-de Boor recursion formula}.
\end{description}
The B-spline basis is typically numerically more stable because the multiple correlation among the basis functions is smaller, but the power spline basis has the advantage that it provides easier interpretation of the knots so as deleting a particular basis function is same as deleting that particular knot. The direct estimation of the regression function $m$ depends on the choice of knot locations and the number of knots. There exist some methods based on the knot-deletion idea. For full details please see Refs.~\refcite{koo91} and \refcite{koo95}.

\subsection{Smoothing Spline}
Consider the following objective function
\begin{align}
\sum_{i=1}^n \{Y_i-m(X_i) \}^2.
\end{align}
Minimizing this function gives the best possible estimate for the unknown regression function. The major problem with the above objective function is that any function $m$ that interpolates the data satisfies it, thereby leading to overfitting. To avoid this, a penalty for the overparametrization is imposed on the function. A convenient way for introducing such a penalty is via the roughness penalty approach. The following function is minimized: 
\begin{align}
\label{ssobj}
\sum_{i=1}^n \{Y_i-m(X_i) \}^2+ \lambda \int \{m"(x) \}^2 dx ,
\end{align}
where $\lambda >0$ is a smoothing parameter. The first term penalizes the lack of fit, which is in some sense modeling bias. The second term denotes the roughness penalty which is related to overparameterization. It is evident that  $\lambda=0$ yields interpolation and $\lambda=\infty$ yields linear regression-typical oversmoothing and undersmoothing types. Hence, the estimate obtained from the objective function $\hat{m}_{\lambda}$, which also depends on the smoothing parameter, is called  the \emph{smoothing spline estimator}. For local properties of the this estimator, please refer to Ref.~\refcite{nyc95}.

It is well known that the solution to the minimization of (\ref{ssobj}) is a cubic spline on the interval $[X_{(1)},X_{(n)}]$ and it is unique in this data range. Moreover, the estimator is a linear smoother with weights that do not depend on the response $\{Y_i \}$\cite{har90}. The connections between kernel regression, which we discussed in the previous section, and smoothing splines have been critically studied by Refs.~\refcite{sil84} and \refcite{sil85}.

The smoothing parameter $\lambda$ can be chosen by minimizing the cross-validation (CV) \cite{sto74,all74} or generalized cross validation (GCV) \cite{wah77,cra78} criteria. Both quantities are consistent estimates of the MISE of $\hat{m}_{\lambda}$. For other methods and details please see Ref.~\refcite{wah90}. Further, for computational issues please refer to Refs.~\refcite{wah90} and \refcite{eub88}.

\section{Additive Models}\label{sec-additive}
While the smoothing methods discussed in the previous section are mostly univariate, the additive model is a widely used multivariate smoothing technique. An additive model is defined as 
\begin{align}
Y&=\alpha+\sum_{j=1}^p f_j(X_j) +\epsilon,
\end{align}
where the errors $\epsilon$ are independent of the $X_j$s and have mean $E(\epsilon)=0$ and variance  $var(\epsilon)=\sigma^2$. The $f_j$ are arbitrary univariate functions, one for each predictor. Since each variable is represented separately, the model retains the interpretative ease of a linear model. 

The most general method for estimating additive models allows us to estimate each function by an arbitrary smoother. Some possible candidates are smoothing splines and kernel smoothers. The backfitting algorithm is a general purpose algorithm that enables one to fit additive models with any kind of smoothing functions although for specific smoothing functions such as smoothing splines or penalized splines there exist separate estimation methods based on least squares. 
The backfitting algorithm is an iterative algorithm and consists of the following steps:
\begin{enumerate}[label=(\roman*)]
\item Initialize: $\alpha=ave(y_i),f_j=f_j^0,j=1,\ldots,p$
\item Cycle:  for $j=1,\ldots,p$ repeat $f_j=S_j\Big(y-\alpha-\sum_{k\neq j}f_k|x_j \Big).$
\item Continue (ii) until the individual functions do not change
\end{enumerate}
where $S_j(y|x_j)$ denotes a smooth of the response $y$ against the predictor $x_j$. The motivation for the backfitting algorithm can be understood using conditional expectation. If the additive model is correct then for any $k$, $E\Big(Y-\alpha-\sum_{j\neq k} f_j(X_j)|X_k \Big)=f_k(X_k)$. This suggests the appropriateness of the backfitting algorithm for computing all the $f_j$. 

Ref.~\refcite{sto85} showed in the context of regression splines - OLS estimation of spline models- that the additive model has the desirable property of reducing a full $p$-dimensional nonparametric regression problem to one that can be fitted with the same asymptotic efficiency as a univariate problem. Due to lack of explicit expressions, the earlier research  by Ref.~\refcite{buj89} studied only the bivariate additive model in detail and showed that both the convergence of the algorithm and uniqueness of its solution depend on the behavior of the product of the two smoothers matrices. Later, Refs.~\refcite{ops00} and \refcite{mam99} extended the convergence theory to $p$-dimensions. For more details, please see Refs.~\refcite{buj89}, \refcite{mam99} and \refcite{ops00}.

We can write all the estimating equations in a compact form:
\begin{align*}
\begin{pmatrix}
I & S_1 &S_1 & \cdots & S_1 \\
S_2 & I &S_2 & \cdots & S_2 \\
\vdots & \vdots & \vdots & \ddots &\vdots \\
S_p & S_p &S_p \cdots & I
\end{pmatrix}
\begin{pmatrix}
f_1\\
f_2\\
\vdots\\
f_p
\end{pmatrix}
&= \begin{pmatrix}
S_1y \\
S_2y\\
\vdots\\
S_py 
\end{pmatrix}\\
\hat{P}f &=\hat{Q}y.
\end{align*}
Backfitting is a Gauss-Seidel procedure for solving the above system. While one could directly use QR decomposition without any iterations to solve the entire system, the computational complexity prohibits doing so. The difficulty is that QR require $O\{(np)^3 \}$ operations while the backfitting involves only $O(np)$ operations, which is much cheaper. For more details see Ref.~\refcite{has90}.

\begin{figure}[H]
\centerline{
\includegraphics[width=8cm]{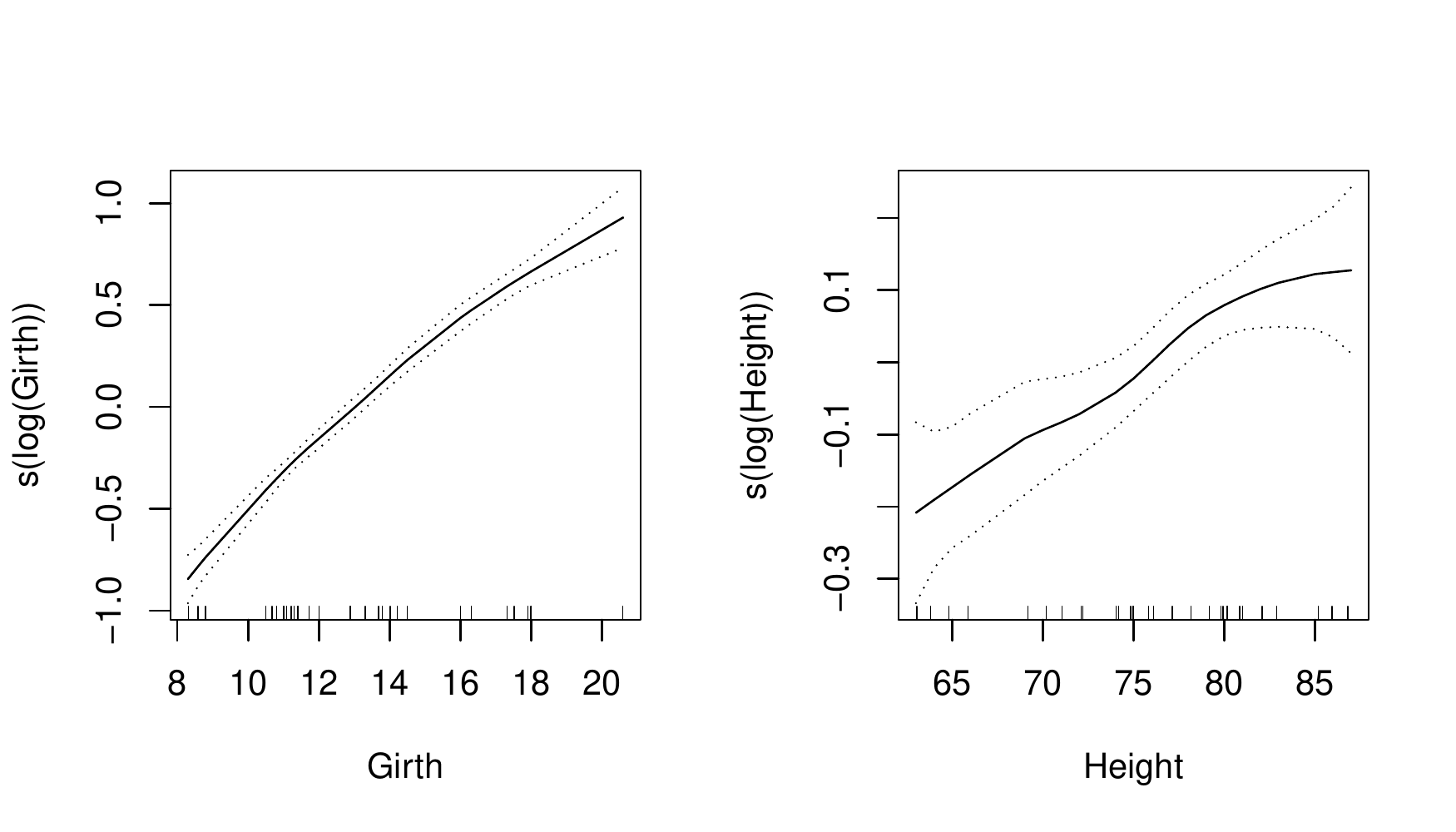}}
\caption{Estimated functions for girth (left) and height (right) using smoothing splines.}
\label{addplot}
\end{figure}

We return to the trees dataset that we considered in the last section, and fit an additive model with both height and girth as predictors to model volume. The fitted model is \emph{log(Volume) $\sim$ log(Height) + log(Girth)}. We used the \emph{gam}\cite{has11} package in R-software. The fitted functions are shown in Figure \ref{addplot}. 

The backfitting method is very generic in the sense that it can handle any type of smoothing function. However, there exists another method specific to penalized splines, which has become quite popular. This method uses penalized splines by estimating the model using penalized regression methods. In equation (\ref{ssobj}), because $m$ is linear in parameters $\beta$, the penalty can always be written as a quadratic form in $\beta$:
\begin{align*}
\int \{ m"(x)\}^2 dx &= \beta^TS\beta,
\end{align*}
where $S$ is the matrix of known coefficients. Therefore the penalized regression spline fitting problem is to minimize
\begin{align}
||y-X\beta||^2+\lambda \beta^TS\beta,
\end{align}
w.r.t $\beta$. It is straightforward to see that the solution is least squares type of estimator 
and depends on the smoothing parameter $\lambda$:
\begin{align}
\hat{\beta} &= (X^TX+\lambda S)^{-1}X^Ty.
\end{align}
Penalized likelihood maximization can only estimate model coefficients $\beta$ given the smoothing parameter $\lambda$. There exist two basic useful estimation approaches: when the scale parameter in the model is known, one can use Mallow's $C_p$ criterion; when the scale parameter is unknown, one can use GCV. Furthermore, for models such as generalized linear models which are estimated iteratively, numerically there exist two different ways of estimating the smoothing parameter:
\begin{description}
\item[Outer iteration:] The score can be minimized directly. This means that the penalized regression must be evaluated for each trial set of smoothing parameters
\item[Performance iteration:]  The score can be minimized and the smoothing parameter selected for each working penalized linear model. This method is computationally efficient.
\end{description}
Performance iteration was originally proposed by Ref.~\refcite{gu92}. It usually converges, and requires only a reliable and efficient method for score minimization. However, it also has some issues related to convergence. In contrast, the outer method suffers from none of the disadvantages that performance iteration has but it is more computationally costly. For more details, please see Ref.~\refcite{woo06}. 

The recent work by Ref.~\refcite{woo15} showcases the successful application of additive models on large datasets and uses performance iteration with block QR updating. This indicates the feasibility of applying these computationally intensive and useful models for big data. The routines are available in the R-package \emph{mgcv}\cite{woo01}.

\section{Markov Chain and Monte Carlo}\label{sec-mcmc}
The Markov Chain Monte Carlo (MCMC) methodology provides enormous scope for realistic and complex statistical modeling. The idea is to perform Monte Carlo integration using Markov chains. Bayesian statisticians, and sometimes also frequentists, need to integrate over possibly high-dimensional probability distributions to draw inference about model parameters or to generate predictions. For a brief history and overview please refer to Refs.~\refcite{gey11} and \refcite{gil05}.

\subsection{Markov chains}
Consider a sequence of random variables, $\{X_0,X_1,X_2,\ldots \}$, such that at each time $t \geq 0$, the next state $X_{t+1}$ is sampled from a conditional distribution $P(X_{t+1}|X_t)$, which depends only on the current state of the chain $X_t$. That is, given the current state $X_t$, the next state $X_{t+1}$ does not depend on the past states - this is called the memory-less property. This sequence is called a Markov chain.

The joint distribution of a Markov chain is determined by two components:
\begin{description}
\item[The marginal distribution of $X_0$], called the initial distribution
\item[The conditional density $p(\cdot|\cdot)$], called the transitional kernel of the chain. 
\end{description}
 It is assumed that the chain is time-homogeneous, which means that the probability $P(\cdot|\cdot)$ does not depend on time $t$. The set in which $X_t$ takes values is called the \emph{state space} of the Markov chain and it can be countably finite or infinite.
 
Under some regularity conditions, the chain will gradually forget its initial state and converge to a unique stationary (invariant) distribution, say $\pi(\cdot)$, which does not depend on $t$ and $X_0$. To converge to a stationary distribution, the chain needs to satisfy three important properties. First, it has to \emph{irreducible}, which means that from all starting points the Markov chain can reach any non empty set with positive probability in some iterations. Second, the chain needs to \emph{aperiodic}, which means that it should not oscillate between any two points in a periodic manner. Third, the chain must be \emph{positive recurrent} as defined next.
\begin{definition}[Ref.~\refcite{myk95}]\\
(i) Markov chain $X$ is called irreducible if for all $i,j$, there exists $t>0$ such that $P_{i,j}(t)=P[X_t=j|X_0=i] >0$. \\ \\
(ii) Let $\tau_{ii}$ be the time of the first return to state $i$, $(\tau_{ii}=min\{t>0:X_t=i|X_0=i\})$. An irreducible chain $X$ is recurrent if $P[\tau_{ii}<\infty]=1$ for some $i$. Otherwise . $X$ is transient. Another equivalent condition for recurrence is
\begin{align*}
\sum_t P_{ij}(t) =\infty,
\end{align*}
for all $i,j$. \\ \\
(iii) An irreducible recurrent chain $X$ is called positive recurrent if $E[\tau_{ii}] < \infty$ for some $i$. Otherwise, it is called null-recurrent. Another equivalent condition for positive recurrence is the existence of a stationary probability distribution for $X$, that is there exists $\pi(\cdot)$ such that 
\begin{align}
\sum_i \pi(i) P_{ij}(t) =\pi(j)
\label{stationary}
\end{align}
for all $j$ and $t \geq 0$. \\ \\
(iv) An irreducible chain $X$ is called a periodic if for some $i$,
greatest common divider $\{t>0:P_{ii}(t) >0 \}=1$.
\end{definition}
In MCMC, since we already have a target distribution $\pi(\cdot)$, then $X$ will be positive recurrent if we can demonstrate irreducibility.

After a sufficiently long burn-in of, say, $m$ iterations, points $\{X_t; t=m+1,\ldots,n\}$ will be the dependent sample approximately from $\pi(\cdot)$. We can now use the output from the Markov chain to estimate the required quantities. For example,  we estimate $E[f(X)]$, where $X$ has distribution $\pi(\cdot)$ as follows:
\begin{align*}
\bar{f} &= \frac{1}{n-m} \sum_{t=m+1}^n f(X_t)
\end{align*}
This quantity is called an \emph{ergodic average} and its convergence to the required expectation is ensured by the ergodic theorem\cite{rob96,myk95}.
\begin{theorem}
 If $X$ is positive recurrent and aperiodic then its stationary distribution $\pi(\cdot)$ is the unique probability distribution satisfying equation (\ref{stationary}). We then say that $X$ is ergodic and the following consequences hold:
 \begin{enumerate}[label=(\roman*)]
\item $P_{ij}(t) \rightarrow \pi(j)$ as $t \rightarrow \infty$ for all $i,j$.
\item (Ergodic theorem) If $E[|f(X)|] <\infty$, then
\begin{align*}
P\Big(\bar{f} \rightarrow E[f(X)]\Big) &=1,
\end{align*}
where $E[f(X)] =\sum_i f(i) \pi(i) $, the expectation of $f(X)$ with respect to $\pi(\cdot)$.
\end{enumerate}
\end{theorem}
Most of the Markov chains procedures in MCMC are reversible which means that they are 
positive recurrent with stationary distribution $\pi(\cdot)$, and $\pi(i) P_{ij} =\pi(j) P_{ji}$.

Further, we say that $X$ is geometrically ergodic, if it is ergodic (positive recurrent and aperiodic) and there exists $0 \leq \lambda <1$ and a function $V(\cdot) >1$ such that
\begin{align}
\sum_j|P_{ij}(t)-\pi(j)| \leq V(i) \lambda^t,
\end{align}
for all $i$. The smallest $\lambda$ for which there exists a function satisfying the above equation is called the rate of convergence. As a consequence to the geometric convergence, the central limit theorem can be used for ergodic averages, that is
\begin{align*}
N^{1/2}(\bar{f}-E[f(X)]) \rightarrow N(0,\sigma^2),
\end{align*}
for some positive constant $\sigma$, as $N\rightarrow \infty$, with the convergence in distribution. For an extensive treatment of geometric convergence and central limit theorems for Markov chains, please refer to  Ref.~\refcite{mey93}.

\subsection{The Gibbs sampler and Metropolis-Hastings algorithm}
Many MCMC algorithms are  hybrids or generalizations of the two simplest methods: the Gibbs sampler and the Metropolis-Hastings algorithm. We therefore describe each of these two methods next.

\subsubsection{Gibbs Sampler}
The Gibbs sampler enjoyed an initial surge of popularity starting with the paper of Ref.~\refcite{gem84} (in a study of image processing models), while the roots of this method can be traced back to Ref.~\refcite{met53}. The Gibbs sampler is a technique for indirectly generating random variables from a (marginal) distribution, without calculating the joint density. With the help of techniques like these we are able to avoid difficult calculations, replacing them with a sequence of easier calculations.

Let $\pi(x)=\pi(x_1,\ldots,x_k),x\in \mathbb{R}^n$ denote a joint density, and let $\pi(x_i|x_{-i})$ denote the induced full conditional densities for each of the components $x_i$, given values of other components $x_{-i}=(x_j;j\neq i), i=1,\ldots,k, 1<k \leq n$. Now the Gibbs sampler proceeds as follows. First, choose arbitrary starting values $x^0=(x_1^0,\ldots,x_k^0)$. Then successively make random drawings from the full conditional distributions $\pi(x_i|x_{-i}),i=1,\ldots,k$ as follows \cite{smi93}:
\begin{center}
\raggedright
$x_1^1$ from $\pi(x_1|x_{-1}^0)$, \\
$x_2^1$ from $\pi(x_2|x_1^1,x_3^0,\ldots,x_k^0)$, \\
$x_3^1$ from $\pi(x_3|x_1^1,x_2^1,x_4^0,\ldots,x_k^0)$, \\
\vdots 
$x_k^1$ from $\pi(x_k|x_{-k}^1)$.
\end{center}
This completes a transition from $x^0=(x_1^0,\ldots,x_k^0)$ to $x^1=(x_1^1,\ldots,x_k^1)$. Each complete cycle through the conditional distributions produces a sequence $x^0,x^1,\ldots,x^t,\ldots$ which is a realization of the Markov chain, with transition probability from $x^t$ to $x^{t+1}$ given by
\begin{align}
TP(x^t,x^{t+1}) &= \Pi_{l=1}^k \pi(x_l^{t+1}|x_j^t,j>l,x_j^{t+1},j<l).
\end{align}
Thus the key feature of this algorithm is to only sample from the full conditional distributions which are often easier to evaluate rather than the joint density. For more details see Ref.~\refcite{cas92}.

\subsubsection{Metropolis-Hastings Algorithm}
The Metropolis-Hastings (M-H) algorithm was developed by Ref.~\refcite{met53}. This algorithm is extremely versatile and produces the Gibbs sampler as a special case\cite{gel93}. 

To construct a Markov chain $X_1,X_2,\ldots,X_t,\ldots$ with state space $\chi$ and equilibrium distribution $\pi(x)$, the Metropolis-Hastings algorithm constructs the transition probability from $X_{t}=x$ to the next realized state $X_{t+1}$ as follows. Let $q(x,x')$ denote a candidate generating density such that 
$X_t=x,x'$ drawn from $q(x,x')$ is considered as a proposed possible value for $X_{t+1}$. With some probability $\alpha(x,x')$, we accept $X_{t+1}=x'$; otherwise, we reject the value generated from $q(x,x')$ and set $X_{t+1}=x$. This construction defines a Markov chain with transition probabilities given as
\begin{align*}
p(x,x') &= \begin{cases}
q(x,x') \alpha(x,x') & \mbox{if} x' \neq x \\
1-\sum_{x''} q(x,x'') \alpha(x,x'') & \mbox{if} x' = x. 
\end{cases}
\end{align*}
Next, we choose
\begin{align*}
\alpha(x,x') &=\begin{cases}
\text{min} \{\frac{\pi(x')q(x',x)}{\pi(x)q(x,x')},1\} & \text{if} \pi(x)q(x,x')>0,\\
1 & \text{if} \pi(x) q(x,x')=0.
\end{cases}
\end{align*}
The choice of the arbitrary $q(x,x')$ to be irreducible and aperiodic is a sufficient condition for $\pi(x)$ to be the equilibrium distribution of the constructed chain.

It can be observed that different choices of $q(x,x')$ will lead to different specific algorithms. For $q(x,x')=q(x',x)$, we have $\alpha(x,x')=min\{\pi(x')/\pi(x),1 \}$, which is the well-known Metropolis algorithm\cite{met53}. For $q(x,x')=q(x'-x)$, the chain is driven by a random walk process. For more choices and their consequences please refer to Ref.~\refcite{tie94}. Similarly for applications of the M-H algorithm and for more details see Ref.~\refcite{chi95}.

\subsection{MCMC Issues}
There is a great deal of theory about the convergence properties of MCMC. However, it has not been found to be very useful in practice for determining the convergence information. A critical issue for users of MCMC is how to determine when to stop the algorithm. Sometimes a Markov chain can appear to have converged to the equilibrium distribution when it has not. This can happen due to the prolonged transition times between state space or due to the multimodality nature of the equilibrium distribution. This phenomenon is often called  \emph{pseudo convergence} or \emph{multimodality}.  

The phenomenon of pseudo convergence has led many MCMC users to embrace the idea of comparing multiple runs of the sampler started at multiple points instead of the usual single run. It is believed that if the multiple runs converge to the same equilibrium distribution then everything is fine with the chain.
However, this approach does not alleviate all the problems. Many times running multiple chains leads to avoiding running the sampler long enough to detect if there are any problems, such as bugs in the code, etc. Those who have used MCMC in complicated problems are probably familiar with stories about last minute problems after running the chain for several weeks. 
In the following we describe the two popular MCMC diagnostic methods. 

Ref.~\refcite{gel92} proposed a convergence diagnostic method which commonly known as "Gelman-Rubin" diagnostic method. It consists of the following two steps. First, obtain an an overdispersed estimate of the target distribution and from it generate the starting points for the desired number of independent chains (say 10). Second, run the Gibbs sampler and re-estimate the target distribution of the required scalar quantity as a conservative Student t distribution, the scale parameter of which involves both the between-chain variance and within-chain variance. Now the convergence is monitored by estimating the factor by which the scale parameter might shrink if sampling were continued indefinitely, namely
\begin{align*}
\sqrt[]{\hat{R}} &= \sqrt[]{\bigg(\frac{n-1}{n}+\frac{m+1}{mn}\frac{B}{W} \bigg) \frac{df}{df-2}},
\end{align*}
where $B$ is the variance between the means from the $m$ parallel chains, $W$ is the within-chain variances,  $df$ is the degrees of freedom of the approximating density, and $n$ is number of observations that are used to reestimate the target density. The authors recommend an iterative process of running additional iterations of the parallel chains and redoing step 2 until the shrink factors for all the quantities of interest are near 1. Though created for the Gibbs sampler, the method by Ref.~\refcite{gel92} may be applied to the output of any MCMC algorithm. It emphasizes the reduction of bias in estimation. There also exist a number of criticisms for the Ref.~\refcite{gel92} method. It relies heavily on the user's ability to find a starting distribution which is highly overdispersed with the target distribution. This means that the user should have some prior knowledge on the target distribution. Although the approach is essentially univariate the authors suggested using -2 times log posterior density as a way of summarizing the convergence of a joint density.

Similarly, Ref.~\refcite{raf92} proposed a diagnostic method which is intended both to detect convergence to the stationary distribution and to provide a way of bounding the variance estimates of the quantiles of functions of parameters. The user must first run a single chain Gibbs sampler with the minimum number of iterations that would be needed to obtain the desired precision of estimation if the samples were independent. The approach is based on the two-state Markov chain theory, as well as the standard sample size formulas that involves formulas of binomial variance. For more details please refer to Ref.~\refcite{raf92}. Critics point out that the method can produce variable estimates of the required number of iterations needed given different initial chains for the same problem and that it is univariate rather than giving information about the full joint posterior distribution. 

There are more methods available to provide the convergence diagnostics for MCMC although not as popular as these. For a discussion about other methods refer to Ref.~\refcite{cow96}.

\begin{figure}[H]
\centerline{
\includegraphics[width=9cm,height=11cm]{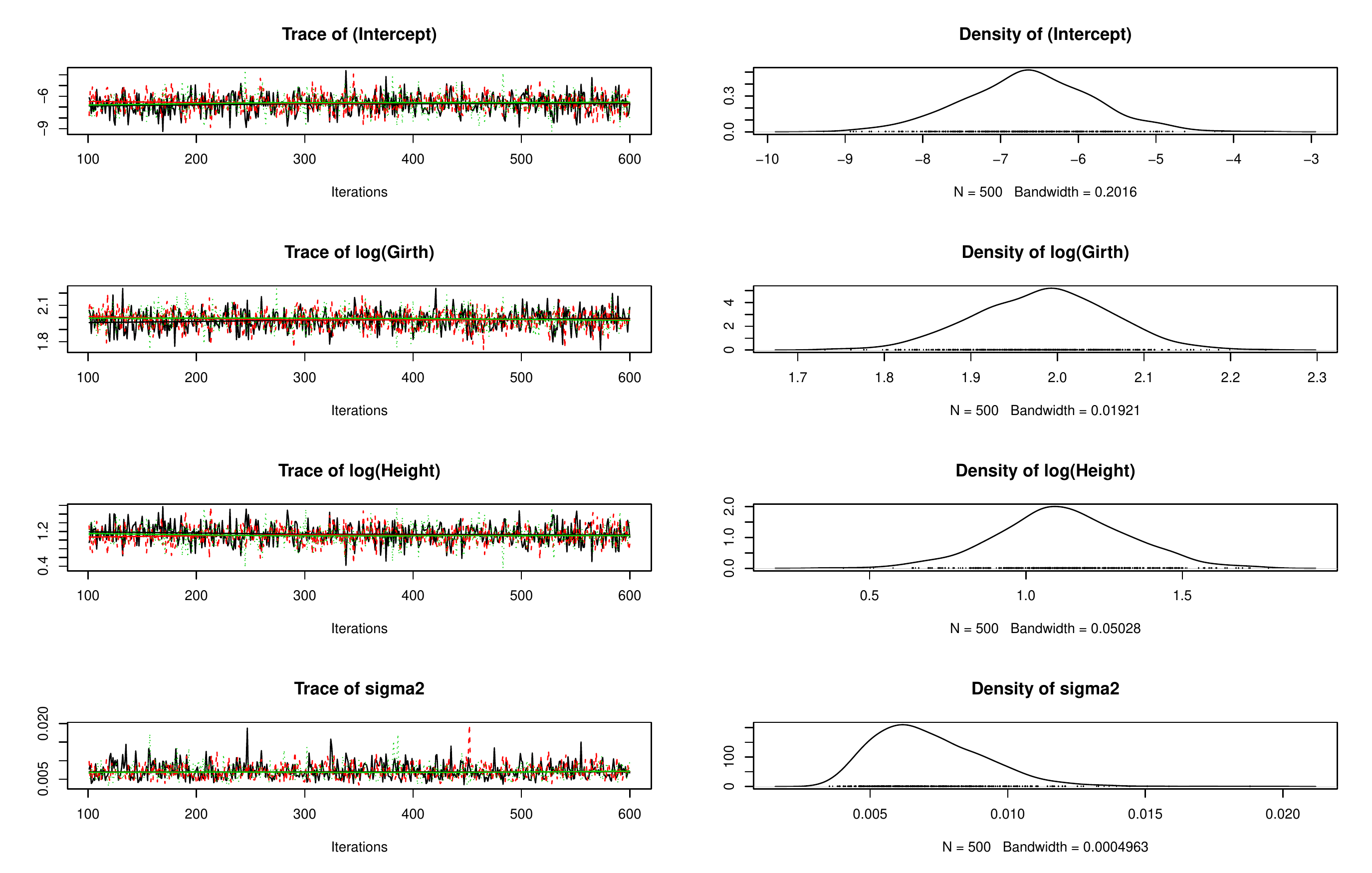}}
\caption{MCMC trace plots with 3 chains for each parameter in the estimated model.}
\label{mcmc}
\end{figure}

\begin{figure}[H]
\centerline{
\includegraphics[width=8cm,height=8cm]{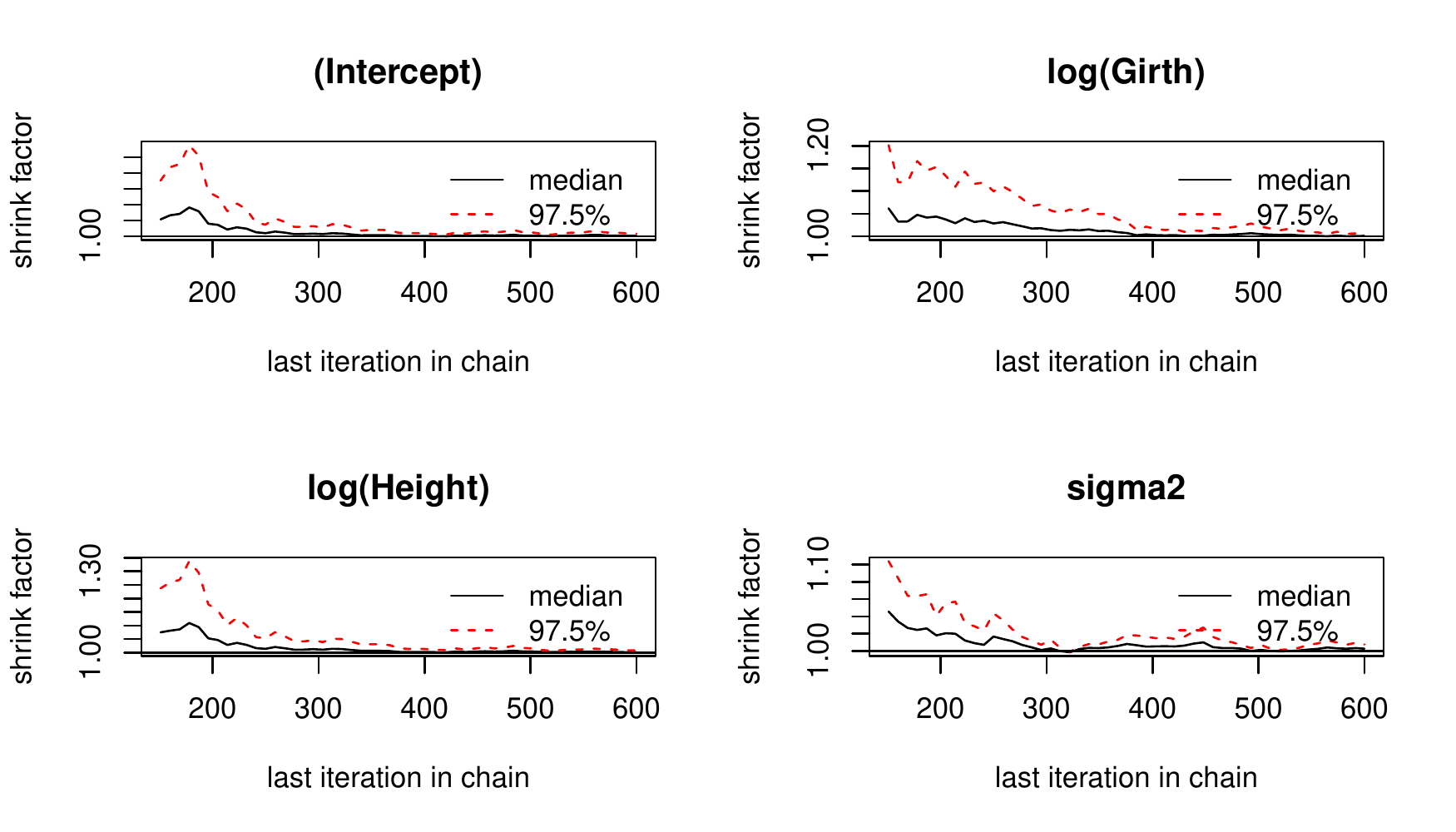}}
\caption{Gelman-Rubin diagnostics for each model parameter, using the results from 3 chains.}
\label{grplot}
\end{figure}

Continuing our illustrations using the trees data, we fit the same model that we used in section \ref{sec-additive} using MCMC methods. We used the \emph{MCMCpack}\cite{mar11} package in R. For the sake of illustration, we considered 3 chains and 600 observations per chain. Among 600 only 100 observations are considered for the burnin. The summary results are described as follows.
{\small 
\begin{verbatim}

Iterations = 101:600
Thinning interval = 1 
Number of chains = 3 
Sample size per chain = 500 

1. Empirical mean and standard deviation for each variable,
   plus standard error of the mean:

                 Mean       SD  Naive SE Time-series SE
(Intercept) -6.640053 0.837601 2.163e-02      2.147e-02
log(Girth)   1.983349 0.078237 2.020e-03      2.017e-03
log(Height)  1.118669 0.213692 5.518e-03      5.512e-03
sigma2       0.007139 0.002037 5.259e-05      6.265e-05

2. Quantiles for each variable:

                 2.5%       25%       50%       75%    97.5%
(Intercept) -8.307100 -7.185067 -6.643978 -6.084797 -4.96591
log(Girth)   1.832276  1.929704  1.986649  2.036207  2.13400
log(Height)  0.681964  0.981868  1.116321  1.256287  1.53511
sigma2       0.004133  0.005643  0.006824  0.008352  0.01161

\end{verbatim}}
Further, the trace plots for the each parameter are displayed in Figure \ref{mcmc}. From the plots it can be observed that the chains are very mixed which gives an indication of convergence of MCMC.
%
Since, we used 3 chains, we can use the Gelman-Rubin convergence diagnostic method and check whether the shrinkage factor is close to 1 or not, which indicates the convergence of 3 chains to the same equilibrium distribution. The results are shown in Figure \ref{grplot}. From the plots we see that for all the parameters the shrink factor and its 97.5\% value are very close to 1, which confirms that the 3 chains converged to the same equilibrium distribution.  

\section{Resampling Methods}\label{sec-resampling}
Resampling methods are statistical procedures that involves repeated sampling of the data. They replace theoretical derivations required for applying traditional methods in statistical analysis by repeatedly resampling the original data and making inference from the resamples. Due to the  advances in computing power these methods have become prominent and particularly well appreciated by applied statisticians. The jacknife and bootstrap are the most popular data-resampling methods used in statistical analysis. For a comprehensive treatment of these methods see Refs.~\refcite{efr82} and \refcite{sha12}. In the following, we describe the Jackknife and bootstrap methods.

\subsection{The Jackknife}
Quenouille\cite{que49} introduced a method, later named the jackknife, to estimate the bias of an estimator by deleting one data point each time from the original dataset and recalculating the estimator based on the rest of the data. Let $T_n=T_n(X_1,\ldots,X_n)$ be an estimator of an unknown parameter $\theta$. The bias of $T_n$ is defined as
\begin{align*}
bias(T_n) &=E(T_n)-\theta.
\end{align*}
Let $T_{n-1,i}=T_{n-1}(X_1,\ldots,X_{i-1},X_{i+1},\ldots,X_n)$ be the given statistic but based on $n-1$ observations $X_1,\ldots,X_{i-1},X_{i+1},\ldots,X_n, i=1,\ldots,n$. Quenouille's jackknife bias estimator is
\begin{align}
b_{JACK} &= (n-1)(\tilde{T}_n-T_n),
\end{align}
where $\tilde{T}_n=n^{-1}\sum_{i=1}^{n}T_{n-1,i}$. This leads to a bias reduced jackknife estimator of $\theta$,
\begin{align}
T_{JACK} &=T_n-b_{JACK} =nT_n-(n-1)\tilde{T}_n.
\end{align}
The jackknife estimators $b_{JACK}$ and $T_{JACK}$ can be heuristically justified as follows. Suppose that
\begin{align}
bias(T_n) &= \frac{a}{n}+\frac{b}{n^2}+O\left(\frac{1}{n^3}\right),
\end{align}
where $a$ and $b$ are unknown but do not depend on $n$. Since $T_{n-1,i},i=1,\ldots,n,$ are identically distributed,
\begin{align}
bias(T_{n-1,i}) &= \frac{a}{n-1}+\frac{b}{(n-1)^2}+O\left(\frac{1}{(n-1)^3}\right),
\end{align}
and $bias(\tilde{T}_n)$ has the same expression. Therefore,
\begin{align*}
E(b_{JACK}) &= (n-1)[bias(\tilde{T}_n)-bias(T_n)] \\
&= (n-1)\bigg[(\frac{1}{n-1}-\frac{1}{n})a + (\frac{1}{(n-1)^2}-\frac{1}{n^2})b +O\left(\frac{1}{n^3}\right) \bigg] \\
&= \frac{a}{n}+\frac{(2n-1)b}{n^2(n-1)} +O\left(\frac{1}{n^2}\right),
\end{align*}
which means that as an estimator of the bias of $T_n$, $b_{JACK}$ is correct up to the order of $n^{-2}$. It follows that 
\begin{align*}
bias(T_{JACK}) =bias(T_n)-E(b_{JACK}) = -\frac{b}{n(n-1)} +O\left(\frac{1}{n^2}\right),
\end{align*}
that is, the bias of $T_{JACK}$ is of order $n^{-2}$. The jackknife produces a bias reduced estimator by removing the first order term in $bias(T_n)$.

The jackknife has become a more valuable tool since Ref.~\refcite{tuk58} found that the jackknife can also be used to construct variance estimators. It is less dependent on model assumptions and does not need any theoretical formula as required by the traditional approach. Although it was prohibitive in the old days due to its computational costs, today it is certainly a popular tool in data analysis.

 \subsection{The Bootstrap}
The bootstrap\cite{efr92} is conceptually the simplest of all  resampling methods. Let $X_1,\ldots,X_n$ denote the dataset of $n$ independent and identically distributed (iid) observations from an unknown distribution $F$ which is estimated by $\hat{F}$, and let $T_n=T_n(X_1,\ldots,X_n)$ be a given statistic. Then the variance of $T_n$ is 
\begin{align}
var(T_n) &= \int \bigg[T_n(x)-\int T_n(y) d \Pi_{i=1}^n F(y_i) \bigg]^2 d \Pi_{i=1}^n F(x_i),
\end{align}
where $x=(x_1,\ldots,x_n)$ and $y=(y_1,\ldots,y_n)$.
Substituting $\hat{F}$ for $F$, we obtain the bootstrap variance estimator
\begin{align*}
\nu_{BOOT} &= \int \bigg[T_n(x)-\int T_n(y) d\Pi_{i=1}^n \hat{F}(y_i)\bigg]^2 d\Pi_{i=1}^n \hat{F}(x_i) \\
&= var_* [T_n(X_1^*,\ldots,X_n^*)|X_1,\ldots,X_n ],
\end{align*}
where $\{X_1^*,\ldots,X_n^* \}$ is an iid sample from $\hat{F}$ and is called a bootstrap sample. $var_*[X_1,\ldots,X_n ]$ denotes the conditional variance for the given $X_1,\ldots,X_n$. The variance cannot be used directly for practical applications when $\nu_{BOOT}$ is not an explicit function of $X_1,\ldots,X_n$. Monte Carlo methods can be used to evaluate this expression when $F$ is known. That is, we repeatedly draw new datasets from $F$ and then use the sample variance of the values $T_n$ computed from new datasets as a numeric approximation to $var(T_n)$. Since $\hat{F}$ is a known distribution, this idea can be further extended. That is, we can draw $\{X_{1b}^{*},\ldots,X_{nb}^{*} \}, b=1\ldots,B$, independently from $\hat{F}$, conditioned on $X_1,\ldots,X_n$. Let $T_{n,b}^{*}=T_n(X_{1b}*,\ldots,X_{nb}^*)$ then we approximate $\nu_{BOOT}$ using the following approximation:
\begin{align}
\nu_{BOOT}^{(B)} &=\frac{1}{B}\sum_{b=1}^B\left(T_{n,b}^*-\frac{1}{B}\sum_{l=1}^B T_{n,l}^* \right)^2.
\label{bootest}
\end{align}
From the law of large numbers, $\nu_{BOOT}=lim_{B\rightarrow \infty}\nu_{BOOT}^{(B)}$ almost surely. Both $\nu_{BOOT}$ and its Monte Carlo approximations $\nu_{BOOT}^{(B)}$ are called bootstrap estimators. While $\nu_{BOOT}^{(B)}$ is more useful for practical applications, $\nu_{BOOT}$ is convenient for theoretical derivations. The distribution $\hat{F}$ used to generate the bootstrap datasets can be any estimator (parametric or nonparametric) of $F$ based on $X_1,\ldots,X_n$. A simple nonparametric estimator of $F$ is the empirical distribution. 
While we have considered the bootstrap variance estimator here, the bootstrap method can be used for more general problems such as inference for regression parameters, hypothesis testing etc,. 
For further discussion of the bootstrap see Ref.~\refcite{efr94}.

Next, we consider the bias and variance of the bootstrap estimator.
Efron Ref.~\refcite{efr92} applied the delta method to approximate the bootstrap bias and variance. Let $\{X_1^*,\ldots,X_n^*\}$ be a bootstrap sample from the empirical distribution $F_n$. 
Define
\begin{align*}
P_i^* &=(\text{the number of} X_j^*=X_i, j=1,\ldots,n)/n ,\\
\text{and}\\
P^* &=(P_1^*,\ldots,P_n^*)'.
\end{align*}
Given $X_1,\ldots,X_n$, the variable $\quad nP^*$ is distributed as a multinomial variable with parameters $n$ and $P_0=(1,\ldots,1)'/n$. Then
\begin{align*}
E_*P^* = P^0 \quad \text{and} \quad var_*(P^*)=n^{-2}(I-\frac{1}{n}11')
\end{align*}
where $I$ is the identity matrix,  $1$ is a column vector of 1's, and $E_*$ and $var_*$ are the bootstrap expectation and variance, respectively.

Now, define a bootstrap estimator of the moment of a random variable $R_n(X_1,\ldots,X_n,F)$. The properties of bootstrap estimators enables us to substitute the population quantities with the empirical quantities $R_n(X_1^*,\ldots,X_n^*,F_n)=R_n(P^*)$. If we expand this around $P^0$ using a multivariate Taylor expansion, we get the desired approximations for the bootstrap bias and variance:
\begin{align*}
b_{BOOT} &= E_*R_n(P^*) \approx R_n(P^0)+\frac{1}{2n^2} tr(V) \\
\nu_{BOOT} &=var_*R_n(P^*) \approx \frac{1}{n^2} U'U
\end{align*}
where $U=\Delta R_n(P^0)$ and $V=\Delta^2R_n(P^0)$.

\subsection{Comparing the Jackknife and the Bootstrap}
In general, the jackknife will be easier to compute if $n$ is less than, say, the 100 or 200 replicates used by the bootstrap for standard error estimation. However, by looking only at the $n$ jackknife samples, the jackknife uses only limited information about the statistic, which means it might be less efficient than the bootstrap. In fact, it turns out that the jackknife can be viewed as a linear approximation to the bootstrap\cite{efr82}. Hence if the statistics are linear then both estimators agree. However, for nonlinear statistics there is a loss of information. Practically speaking, the accuracy of the jackknife estimate of standard error depends on how close the estimate is to linearity. Also, while it is not obvious how to estimate the entire sampling distribution of $T_n$ by jackknifing, the bootstrap can be readily used to obtain a distribution estimator for $T_n$.

In considering the merits or demerits of the bootstrap, it is to be noted that the general formulas for estimating standard errors that involve the observed Fisher information matrix are essentially bootstrap estimates carried out in a parametric framework. While the use of the Fisher information matrix involves  parametric assumptions, the bootstrap is free of those. The data analyst is free to obtain standard errors for enormously complicated estimators subject only to the constraints of computer time. In addition, if needed, one could obtain more smoothed bootstrap estimates by convoluting the nonparametric bootstrap with the parametric bootstrap- a parametric bootstrap involves generating samples based on the estimated parameters while nonparametric bootstrap involves generating samples based on available data alone.

To provide a simple illustration, we again considered the trees data and fit an ordinary regression model with the formula mentioned in Section \ref{sec-additive}. To conduct a bootstrap analysis on the regression parameters we resampled the data with replacement 100 times (bootstrap replications) and fit the same model to each sample. We calculated the mean and standard deviation for each regression coefficient, which are analogous to OLS coefficient and standard error. We performed the same for the jackknife estimators. The results are produced in Table \ref{bootjack}. From the results it can be seen that the jackknife is off due to the small sample size. However, the bootstrap results are much closer to the values from OLS.
\begin{table}[H]
\tbl{Comparison of the bootstrap, jackknife, and parametric method (OLS) in a regression setting.}
{\begin{tabular}{l c c c }
\hline
 & OLS & Bootstrap & Jackknife\\
\hline
(Intercept) & $-6.63$ & $-6.65$ & $-6.63$ \\
            & $(0.80)$ & $(0.73)$ & $(0.15)$     \\
log(Height) & $1.12$  & $1.12$ & $1.11$\\
            & $(0.20)$ & $(0.19)$ & $(0.04)$     \\
log(Girth)  & $1.98$  & $1.99$ & $1.98$\\
            & $(0.08)$ & $(0.07)$  & $(0.01)$    \\
\hline
Observations   & $31$  & &          \\
Samples        &    & $100$ & $31$       \\
\hline
\end{tabular}}
\label{bootjack}
\end{table}



\section{Conclusion}
The area and methods of computational statistics have been evolving rapidly. Existing statistical software such as R already have efficient routines to implement and evaluate these methods. In addition, there exists literature on parallelizing these methods to make them even more efficient, for e.g., please see Ref.~\refcite{woo15}.

While some of the existing methods are still prohibitive even with moderately large data - such as the local linear estimator - implementations using more resourceful environments such as servers or clouds make such methods feasible even with big data. For an example see Ref.~\refcite{zha13} where they used server (32GB RAM) to estimate their proposed model on the real data which did not take more than 34 seconds. Otherwise, it would have taken more time. This will indicate the helpfulness of the computing power while estimating these computationally intensive methods.

To the best of our knowledge, there exist multiple algorithms or R-packages to implement all the methods discussed here. However, it should be noted that not every method is computationally efficient. For example, Ref.~\refcite{den11} reported that within R software there are 20 packages that implement density estimation. Further, they found that two packages (\emph{KernSmooth, ASH}) are very fast, accurate and also well-maintained. Hence the user should be wise enough to choose efficient implementations when dealing with larger datasets.

Lastly, as we mentioned before, we are able to cover only very few of the modern statistical computing methods. For an expanded exposition of computational methods especially for inference, see Ref.~\refcite{efr16}.


\bibliographystyle{amsplain}
\bibliography{datasc}

\providecommand{\bysame}{\leavevmode\hbox to3em{\hrulefill}\thinspace}
\providecommand{\MR}{\relax\ifhmode\unskip\space\fi MR }
\providecommand{\MRhref}[2]{%
  \href{http://www.ams.org/mathscinet-getitem?mr=#1}{#2}
}
\providecommand{\href}[2]{#2}
\begin{thebibliography}{10}

\bibitem{all74}
David~M Allen, \emph{The relationship between variable selection and data
  agumentation and a method for prediction}, Technometrics \textbf{16} (1974),
  no.~1, 125--127.

\bibitem{atk85}
Anthony~Curtis Atkinson, \emph{Plots, transformations, and regression: an
  introduction to graphical methods of diagnostic regression analysis}, no.
  519.536 A875, Clarendon Press, 1985.

\bibitem{bow84}
Adrian~W Bowman, \emph{An alternative method of cross-validation for the
  smoothing of density estimates}, Biometrika \textbf{71} (1984), no.~2,
  353--360.

\bibitem{bow84cross}
Adrian~W Bowman, Peter Hall, and DM~Titterington, \emph{Cross-validation in
  nonparametric estimation of probabilities and probability densities},
  Biometrika \textbf{71} (1984), no.~2, 341--351.

\bibitem{buj89}
Andreas Buja, Trevor Hastie, and Robert Tibshirani, \emph{Linear smoothers and
  additive models}, The Annals of Statistics (1989), 453--510.

\bibitem{cas92}
George Casella and Edward~I George, \emph{Explaining the gibbs sampler}, The
  American Statistician \textbf{46} (1992), no.~3, 167--174.

\bibitem{chi95}
Siddhartha Chib and Edward Greenberg, \emph{Understanding the
  metropolis-hastings algorithm}, The american statistician \textbf{49} (1995),
  no.~4, 327--335.

\bibitem{chu91}
C-K Chu and JS~Marron, \emph{Choosing a kernel regression estimator},
  Statistical Science (1991), 404--419.

\bibitem{cle79}
William~S Cleveland, \emph{Robust locally weighted regression and smoothing
  scatterplots}, Journal of the American statistical association \textbf{74}
  (1979), no.~368, 829--836.

\bibitem{cow96}
Mary~Kathryn Cowles and Bradley~P Carlin, \emph{Markov chain monte carlo
  convergence diagnostics: a comparative review}, Journal of the American
  Statistical Association \textbf{91} (1996), no.~434, 883--904.

\bibitem{cra78}
Peter Craven and Grace Wahba, \emph{Smoothing noisy data with spline
  functions}, Numerische Mathematik \textbf{31} (1978), no.~4, 377--403.

\bibitem{den11}
Henry Deng and Hadley Wickham, \emph{Density estimation in r},  (2011).

\bibitem{efr92}
Bradley Efron, \emph{Bootstrap methods: another look at the jackknife},
  Breakthroughs in Statistics, Springer, 1992, pp.~569--593.

\bibitem{efr82}
Bradley Efron and B~Efron, \emph{The jackknife, the bootstrap and other
  resampling plans}, vol.~38, SIAM, 1982.

\bibitem{efr16}
Bradley Efron and Trevor Hastie, \emph{Computer age statistical inference},
  vol.~5, Cambridge University Press, 2016.

\bibitem{efr94}
Bradley Efron and Robert~J Tibshirani, \emph{An introduction to the bootstrap},
  CRC press, 1994.

\bibitem{epa69}
VA~Epanechnikov, \emph{Nonparametric estimation of a multidimensional
  probability density}, Teoriya veroyatnostei i ee primeneniya \textbf{14}
  (1969), no.~1, 156--161.

\bibitem{eub88}
Randall~L Eubank, \emph{Spline smoothing and nonparametric regression}, no. 04;
  QA278. 2, E8., 1988.

\bibitem{fan92}
Jianqing Fan, \emph{Design-adaptive nonparametric regression}, Journal of the
  American statistical Association \textbf{87} (1992), no.~420, 998--1004.

\bibitem{fan93}
\bysame, \emph{Local linear regression smoothers and their minimax
  efficiencies}, The Annals of Statistics (1993), 196--216.

\bibitem{fan96}
Jianqing Fan and Irene Gijbels, \emph{Local polynomial modelling and its
  applications: monographs on statistics and applied probability 66}, vol.~66,
  CRC Press, 1996.

\bibitem{fan94}
Jianqing Fan and James~S Marron, \emph{Fast implementations of nonparametric
  curve estimators}, Journal of Computational and Graphical Statistics
  \textbf{3} (1994), no.~1, 35--56.

\bibitem{fri98}
Jerome~H Friedman, \emph{Data mining and statistics: What's the connection?},
  Computing Science and Statistics \textbf{29} (1998), no.~1, 3--9.

\bibitem{gas79}
Theo Gasser and Hans-Georg M{\"u}ller, \emph{Kernel estimation of regression
  functions}, Smoothing techniques for curve estimation, Springer, 1979,
  pp.~23--68.

\bibitem{gel93}
Andrew Gelman, \emph{Iterative and non-iterative simulation algorithms},
  Computing science and statistics (1993), 433--433.

\bibitem{gel92}
Andrew Gelman and Donald~B Rubin, \emph{Inference from iterative simulation
  using multiple sequences}, Statistical science (1992), 457--472.

\bibitem{gem84}
Stuart Geman and Donald Geman, \emph{Stochastic relaxation, gibbs
  distributions, and the bayesian restoration of images}, IEEE Transactions on
  pattern analysis and machine intelligence (1984), no.~6, 721--741.

\bibitem{gey11}
Charles Geyer, \emph{Introduction to markov chain monte carlo}, Handbook of
  Markov Chain Monte Carlo (2011), 3--48.

\bibitem{gil05}
Walter~R Gilks, \emph{Markov chain monte carlo}, Wiley Online Library.

\bibitem{gre94}
PJ~Green and BW~Silverman, \emph{Nonparametric regression and generalized
  linear models, vol. 58 of}, Monographs on Statistics and Applied Probability
  (1994).

\bibitem{gu92}
Chong Gu, \emph{Cross-validating non-gaussian data}, Journal of Computational
  and Graphical Statistics \textbf{1} (1992), no.~2, 169--179.

\bibitem{hal83}
Peter Hall, \emph{Large sample optimality of least squares cross-validation in
  density estimation}, The Annals of Statistics (1983), 1156--1174.

\bibitem{har90}
Wolfgang Hardle, \emph{Applied nonparametric regression}, Cambridge, UK (1990).

\bibitem{har12}
Wolfgang H{\"a}rdle, \emph{Smoothing techniques: with implementation in s},
  Springer Science \& Business Media, 2012.

\bibitem{has11}
T~Hastie, \emph{Gam: generalized additive models. r package version 1.06. 2},
  2011.

\bibitem{has93}
Trevor Hastie and Clive Loader, \emph{Local regression: Automatic kernel
  carpentry}, Statistical Science (1993), 120--129.

\bibitem{has90}
Trevor~J Hastie and Robert~J Tibshirani, \emph{Generalized additive models},
  vol.~43, CRC Press, 1990.

\bibitem{hod56}
JOSEPH~L Hodges~Jr and ERIC~L Lehmann, \emph{The efficiency of some
  nonparametric competitors of the t-test}, The Annals of Mathematical
  Statistics (1956), 324--335.

\bibitem{jon96}
M~Chris Jones, James~S Marron, and Simon~J Sheather, \emph{A brief survey of
  bandwidth selection for density estimation}, Journal of the American
  Statistical Association \textbf{91} (1996), no.~433, 401--407.

\bibitem{kat79}
V~Ya Katkovnik, \emph{Linear and nonlinear methods of nonparametric regression
  analysis}, Soviet Automatic Control \textbf{5} (1979), 25--34.

\bibitem{koo91}
Charles Kooperberg and Charles~J Stone, \emph{A study of logspline density
  estimation}, Computational Statistics \& Data Analysis \textbf{12} (1991),
  no.~3, 327--347.

\bibitem{koo95}
Charles Kooperberg, Charles~J Stone, and Young~K Truong, \emph{Hazard
  regression}, Journal of the American Statistical Association \textbf{90}
  (1995), no.~429, 78--94.

\bibitem{loa06}
Clive Loader, \emph{Local regression and likelihood}, Springer Science \&
  Business Media, 2006.

\bibitem{mac89}
YP~Mack and Hans-Georg M{\"u}ller, \emph{Derivative estimation in nonparametric
  regression with random predictor variable}, Sankhy{\=a}: The Indian Journal
  of Statistics, Series A (1989), 59--72.

\bibitem{mam99}
Enno Mammen, Oliver Linton, J~Nielsen, et~al., \emph{The existence and
  asymptotic properties of a backfitting projection algorithm under weak
  conditions}, The Annals of Statistics \textbf{27} (1999), no.~5, 1443--1490.

\bibitem{mar11}
Andrew~D Martin, Kevin~M Quinn, Jong~Hee Park, et~al., \emph{Mcmcpack: Markov
  chain monte carlo in r}, Journal of Statistical Software \textbf{42} (2011),
  no.~9, 1--21.

\bibitem{met53}
Nicholas Metropolis, Arianna~W Rosenbluth, Marshall~N Rosenbluth, Augusta~H
  Teller, and Edward Teller, \emph{Equation of state calculations by fast
  computing machines}, The journal of chemical physics \textbf{21} (1953),
  no.~6, 1087--1092.

\bibitem{mey93}
Sean~P Meyn and Richard~L Tweedie, \emph{Stability of markovian processes ii:
  Continuous-time processes and sampled chains}, Advances in Applied
  Probability (1993), 487--517.

\bibitem{myk95}
Per Mykland, Luke Tierney, and Bin Yu, \emph{Regeneration in markov chain
  samplers}, Journal of the American Statistical Association \textbf{90}
  (1995), no.~429, 233--241.

\bibitem{nad64}
Elizbar~A Nadaraya, \emph{On estimating regression}, Theory of Probability \&
  Its Applications \textbf{9} (1964), no.~1, 141--142.

\bibitem{nyc95}
Douglas Nychka, \emph{Splines as local smoothers}, The Annals of Statistics
  (1995), 1175--1197.

\bibitem{ops00}
Jean~D Opsomer, \emph{Asymptotic properties of backfitting estimators}, Journal
  of Multivariate Analysis \textbf{73} (2000), no.~2, 166--179.

\bibitem{pur62}
E~Purzen, \emph{On estimation of a probability density and mode}, Ann. Math.
  Statist \textbf{39} (1962), 1065--1076.

\bibitem{que49}
Maurice~H Quenouille, \emph{Approximate tests of correlation in time-series 3},
  Mathematical Proceedings of the Cambridge Philosophical Society, vol.~45,
  Cambridge Univ Press, 1949, pp.~483--484.

\bibitem{raf92}
Adrian~E Raftery, Steven Lewis, et~al., \emph{How many iterations in the gibbs
  sampler}, Bayesian statistics \textbf{4} (1992), no.~2, 763--773.

\bibitem{rob96}
Gareth~O Roberts, \emph{Markov chain concepts related to sampling algorithms},
  Markov chain Monte Carlo in practice \textbf{57} (1996).

\bibitem{ros56}
Murray Rosenblatt et~al., \emph{Remarks on some nonparametric estimates of a
  density function}, The Annals of Mathematical Statistics \textbf{27} (1956),
  no.~3, 832--837.

\bibitem{rud82}
Mats Rudemo, \emph{Empirical choice of histograms and kernel density
  estimators}, Scandinavian Journal of Statistics (1982), 65--78.

\bibitem{rup94}
David Ruppert and Matthew~P Wand, \emph{Multivariate locally weighted least
  squares regression}, The annals of statistics (1994), 1346--1370.

\bibitem{rya76}
Thomas A Thomas~A Ryan, Brian~L Joiner, and Barbara~F Ryan, \emph{Minitab
  student handbook}, no. 04; LB1028. 5, R8., 1976.

\bibitem{sha12}
Jun Shao and Dongsheng Tu, \emph{The jackknife and bootstrap}, Springer Science
  \& Business Media, 2012.

\bibitem{sil84}
Bernard~W Silverman, \emph{Spline smoothing: the equivalent variable kernel
  method}, The Annals of Statistics (1984), 898--916.

\bibitem{sil86}
\bysame, \emph{Density estimation for statistics and data analysis}, vol.~26,
  CRC press, 1986.

\bibitem{sil85}
Bernhard~W Silverman, \emph{Some aspects of the spline smoothing approach to
  non-parametric regression curve fitting}, Journal of the Royal Statistical
  Society. Series B (Methodological) (1985), 1--52.

\bibitem{sim12}
Jeffrey~S Simonoff, \emph{Smoothing methods in statistics}, Springer Science \&
  Business Media, 2012.

\bibitem{smi93}
Adrian~FM Smith and Gareth~O Roberts, \emph{Bayesian computation via the gibbs
  sampler and related markov chain monte carlo methods}, Journal of the Royal
  Statistical Society. Series B (Methodological) (1993), 3--23.

\bibitem{sto77}
Charles~J Stone, \emph{Consistent nonparametric regression}, The annals of
  statistics (1977), 595--620.

\bibitem{sto80}
\bysame, \emph{Optimal rates of convergence for nonparametric estimators}, The
  annals of Statistics (1980), 1348--1360.

\bibitem{sto84}
\bysame, \emph{An asymptotically optimal window selection rule for kernel
  density estimates}, The Annals of Statistics (1984), 1285--1297.

\bibitem{sto85}
\bysame, \emph{Additive regression and other nonparametric models}, The annals
  of Statistics (1985), 689--705.

\bibitem{sto74}
Mervyn Stone, \emph{Cross-validatory choice and assessment of statistical
  predictions}, Journal of the Royal Statistical Society. Series B
  (Methodological) (1974), 111--147.

\bibitem{stu73}
Alan Stuart, Maurice~G Kendall, et~al., \emph{The advanced theory of
  statistics}, vol.~2, Charles Griffin, 1973.

\bibitem{tea13}
R~Core Team et~al., \emph{R: A language and environment for statistical
  computing},  (2013).

\bibitem{tie94}
Luke Tierney, \emph{Markov chains for exploring posterior distributions}, the
  Annals of Statistics (1994), 1701--1728.

\bibitem{tuk58}
John~W Tukey, \emph{Bias and confidence in not-quite large samples}, Annals of
  Mathematical Statistics, vol.~29, INST MATHEMATICAL STATISTICS IMS BUSINESS
  OFFICE-SUITE 7, 3401 INVESTMENT BLVD, HAYWARD, CA 94545, 1958, pp.~614--614.

\bibitem{tuk62}
\bysame, \emph{The future of data analysis}, The Annals of Mathematical
  Statistics \textbf{33} (1962), no.~1, 1--67.

\bibitem{wah77}
Grace Wahba, \emph{Practical approximate solutions to linear operator equations
  when the data are noisy}, SIAM Journal on Numerical Analysis \textbf{14}
  (1977), no.~4, 651--667.

\bibitem{wah90}
\bysame, \emph{Spline models for observational data}, vol.~59, Siam, 1990.

\bibitem{wan15}
MP~Wand and BD~Ripley, \emph{Kernsmooth: Functions for kernel smoothing for
  wand and jones (1995). r package version 2.23-15}, 2015.

\bibitem{wat64}
Geoffrey~S Watson, \emph{Smooth regression analysis}, Sankhy{\=a}: The Indian
  Journal of Statistics, Series A (1964), 359--372.

\bibitem{woo06}
Simon Wood, \emph{Generalized additive models: an introduction with r}, CRC
  press, 2006.

\bibitem{woo01}
Simon~N Wood, \emph{mgcv: Gams and generalized ridge regression for r}, R news
  \textbf{1} (2001), no.~2, 20--25.

\bibitem{woo15}
Simon~N Wood, Yannig Goude, and Simon Shaw, \emph{Generalized additive models
  for large data sets}, Journal of the Royal Statistical Society: Series C
  (Applied Statistics) \textbf{64} (2015), no.~1, 139--155.

\bibitem{zha13}
Xiaoke Zhang, Byeong~U Park, and Jane-ling Wang, \emph{Time-varying additive
  models for longitudinal data}, Journal of the American Statistical
  Association \textbf{108} (2013), no.~503, 983--998.

\end{thebibliography}

\end{multicols}
\end{document}